\documentclass[12pt]{article}
\pdfoutput=1
\usepackage{jheppub}

\usepackage{amsmath,amssymb,amsfonts}
\usepackage{graphicx}
\usepackage{color}
\usepackage{tikz}
\usepackage{dsfont}

\usepackage{booktabs}
\usepackage{multirow}

\usepackage{subfigure}

\def\cA{{\cal A}}
\def\cB{{\cal B}}
\def\cC{{\cal C}}

\def\cG{{\cal G}}
\def\cH{{\cal H}}
\def\cM{{\cal M}}

\def\cS{{\cal S}}
\def\cY{{\cal Y}}

\def\q{{\mathfrak q}}

\def\RR{{\mathds{R}}}
\def\ZZ{{\mathds{Z}}}

\DeclareMathOperator{\vol}{vol}

\makeatletter\def\l@subsubsection#1#2{}%
\makeatother

\def\Im{\mathop{\rm Im}}
\def\Re{\mathop{\rm Re}}

\def\be{\begin{eqnarray}}
\def\ee{\end{eqnarray}}

\newcommand{\pslash}{\ensuremath\diagup\!\!\!\!\!{+}}

\begin{document}

\title{Testing \texorpdfstring{$AdS_6/CFT_5$}{AdS6/CFT5} in Type IIB with stringy operators}

\author[a]{Oren Bergman,}
\emailAdd{bergman@physics.technion.ac.il}
\author[b]{Diego Rodr\'iguez-G\'omez,}
\emailAdd{d.rodriguez.gomez@uniovi.es}
\author[c]{Christoph F.~Uhlemann} 
\emailAdd{uhlemann@physics.ucla.edu}

\affiliation[a]{Department of Physics, Technion, Israel Institute of Technology\\
Haifa, 32000, Israel\\[-4mm]}

\affiliation[b]{Department of Physics, Universidad de Oviedo\\
Avda.~Calvo Sotelo 18, 33007, Oviedo, Spain\\[-4mm]
}

\affiliation[c]{Mani L.\ Bhaumik Institute for Theoretical Physics\\
Department of Physics and Astronomy\\
University of California, Los Angeles, CA 90095, USA\\[-4mm]}

\abstract{
We provide further evidence that the recently constructed warped $AdS_6$ solutions in Type IIB supergravity are dual to 5d SCFTs that correspond to $(p,q)$ 5-brane webs with large numbers of like-charged external 5-branes. 
We study a number of specific examples, including the $T_N$ theory, and identify the bulk states dual to a class of operators with
${\cal O}(N)$ scaling dimensions in terms of strings and string-webs embedded in the solutions.
}

\maketitle

\section{Introduction and summary}

Theories in dimensions greater than four have become an important ingredient in the modern approach to supersymmetric quantum field theories. On the one hand, they provide the basic building blocks, which, upon appropriate compactification, allow to understand virtually all other lower dimensional theories. On the other hand, they often exhibit exotic behavior such as global symmetries of exceptional type. However, defining interacting quantum field theories in dimensions greater than four is challenging for perturbative quantization methods based on a classical Lagrangian. In particular, the coupling ``constant'' in Yang-Mills type gauge theories is dimensionful and the theories are non-renormalizable by power counting. Nevertheless, a large body of evidence, based on field theory arguments and string theory constructions, suggests that such theories can exist as well-defined quantum field theories, making them an intriguing laboratory for exploring quantum field theory beyond perturbation theory. Specifically in $5d$, large classes of supersymmetric gauge theories with eight supercharges are believed to flow to well-defined UV fixed points, with the strongly-coupled superconformal field theories (SCFTs) at the fixed points realizing the unique $5d$ superconformal algebra $F(4)$ \cite{Seiberg:1996bd,Intriligator:1997pq}. A particularly general and versatile approach to engineering these 5d SCFTs and their gauge theory deformations is via 5-brane webs in Type IIB string theory \cite{Aharony:1997ju,Aharony:1997bh}. It allows to construct large classes of SCFTs, some of which allow relevant deformations that flow to gauge theories with various types of gauge groups and matter fields, as well as other SCFTs with no gauge theory deformations at all. 

In the absence of a conventional Lagrangian description, AdS/CFT dualities can provide a particularly valuable tool for quantitative studies of the 5d SCFTs -- if supergravity solutions corresponding to the string theory constructions are available and permit a clear identification with dual 5d SCFTs. A locally unique solution to massive Type IIA supergravity, corresponding to the construction of a particular class of gauge theories in Type I' string theory, has been known for some time \cite{Brandhuber:1999np,Passias:2012vp}, and along with its orbifolds \cite{Bergman:2012kr} has featured in numerous holographic analyses \cite{Jafferis:2012iv,Bergman:2012qh,Assel:2012nf,Bergman:2013koa,Pini:2014bea,Passias:2018swc}. Solutions directly in Type IIB supergravity, on the other hand, have only been constructed rather  recently \cite{DHoker:2016ujz,DHoker:2016ysh,DHoker:2017mds,DHoker:2017zwj}.\footnote{Previous studies of the BPS equations in Type IIB can be found in \cite{Apruzzi:2014qva,Kim:2015hya,Kim:2016rhs}, while T-duals of the Type IIA solution have been discussed in \cite{Lozano:2012au,Lozano:2013oma}.} This includes a large class of physically regular solutions which naturally relate to pure 5-brane webs \cite{DHoker:2017mds}, as well as an extension realizing additional 7-branes in the faces of 5-brane webs \cite{DHoker:2017zwj}.
These physically regular solutions can account for large classes of 5-brane webs and the corresponding field theories that can be engineered in Type IIB string theory. They are characterized explicitly by the types and charges of 5- and 7-branes used in the Type IIB string theory constructions for 5d SCFTs, which allows for a clear identification of the supergravity solutions with corresponding 5-brane webs and 5d SCFTs.
The sphere partition functions of the dual SCFTs were studied holographically in \cite{Gutperle:2017tjo,Gutperle:2018vdd} and used for various consistency checks supporting the identification with 5-brane webs. But independent field theory results for a direct quantitative comparison were not available.

The aim of this work is to provide a quantitative match between AdS/CFT computations using the Type IIB supergravity solutions and results obtained independently from their proposed field theory duals. We discuss several explicit examples of 5-brane webs of the type that the supergravity solutions are identified with, and the field theories that they engineer. 
Within these theories we identify a special class of chiral operators that correspond to strings and string-webs connected to the
external legs of the 5-brane webs. 
These operators have ${\cal O}(N)$ scaling dimensions in the large $N$ limit.
This information can then be compared to the supergravity picture, where we can identify the holographic duals of these operators as states described by strings and string-webs embedded in the Type IIB supergravity background.
Their masses are related via the AdS/CFT dictionary to the scaling dimensions of the dual operators.
In the remainder of this section we will give a more detailed overview and summary of our results. The explicit computations are presented in the main part of the paper.

\subsection{Summary}

The 5-brane webs that are identified with the solutions of \cite{DHoker:2017mds,DHoker:2017zwj} are characterized by large numbers of like-charged external $(p,q)$ 5-branes. Generically, the external 5-branes of a given brane web can be terminated on appropriate 7-branes without breaking supersymmetry \cite{DeWolfe:1999hj}, and if each 5-brane is terminated on a distinct 7-brane, the field theory remains unchanged. For brane webs with multiple 5-branes of equal charge, however, each group of like-charged 5-branes may also be partitioned into subgroups which terminate on the same 7-brane, and due to the $s$-rule different partitions lead to different SCFTs \cite{Benini:2009gi}. For the brane webs that were proposed to correspond to the solutions of \cite{DHoker:2017mds,DHoker:2017zwj}, each external 5-brane is terminated on a distinct 7-brane, which is equivalent to not terminating the external 5-branes on 7-branes at all.
This leads to the maximal global symmetry among the choices to terminate like-charged 5-branes on 7-branes, and in that sense to maximally symmetric 5d SCFTs. 
We denote the charges of the external 5-branes by mutually-prime pairs of integers $(p_i,q_i)$, with $i=1,\ldots,n$, and the numbers of external 5-branes within each group by $N_i$.
Charge conservation requires  $\sum N_i p_i = \sum N_i q_i = 0$.
The global symmetry is in general given by 
\begin{align}\label{eq:global-symmetry}
 \prod_{i=1}^n SU(N_i)\times U(1)^{n-3} \,,
\end{align}
where the $U(1)$ factors are associated to mass deformations of the 5d SCFT that correspond to motion of the external 5-branes in groups.
For some special cases with small values of the $N_i$ the symmetry is larger.

The general strategy for obtaining information on the spectrum of the 5d SCFTs will be to consider deformations that lead to IR quiver gauge theories. 
The latter have a conventional Lagrangian description, and a subset of the operators in the SCFT can be constructed from the Lagrangian fields in the gauge theory. This yields operators in representations of the global symmetry group of the gauge theory, which in general is a subgroup of the global symmetry group of the UV fixed point. 
The scaling dimensions of these operators are protected by the BPS shortening conditions,
and can therefore be extrapolated along the RG flow to the UV SCFT.

The specific set of examples we study is summarized in Table \ref{tab:CFT1}.
This includes the 5d $T_N$ theory, originally introduced in  \cite{Benini:2009gi},
and a theory that we call $+_{N,M}$ corresponding to the intersection of D5-branes and NS5-branes, originally introduced in \cite{Aharony:1997bh}.
We have named the theories according to the shape of the corresponding 5-brane web.
The other examples include the $Y_N$ theory which corresponds to a junction of $(1,1)$, $(-1,1)$ and NS5-branes, the $+_{N,M,k}$ theory which corresponds to a 5-brane web with additional 7-branes inside a face of the web, the $X_{N,M}$ theory which corresponds to an intersection of $(1,1)$ and $(1,-1)$ 5-branes, and the $\pslash_N$ theory which corresponds to an intersection of D5, NS5 and $(1,1)$ 5-branes. 

\begin{table}
\centering
\begin{tabular}{@{\extracolsep{5pt}}cccc}
\toprule   
&global symmetry & stringy operators & $\Delta$ \\
\hline
\hline
$T_N$ & $SU(N)^3$ & $({\bf N},{\bf N},{\bf N})$ & $\frac{3}{2}(N-1)$\\[1mm]
\hline
$Y_N$ & $SU(2N)\,{\times}\, SU(N)^2$ & $(({\bf 2N})^2_{asym},{\bf N},{\bf N})$ & $3(N-1)$\\[1mm]
\hline
$+_{N,M}$ & $SU(N)^2\,{\times}\, SU(M)^2\,{\times}\, U(1)$ & $({\bf N},{\bf \bar N},{\bf 1},{\bf 1})_{M}$ & $\frac{3}{2}M$\\
&& $({\bf 1},{\bf 1},{\bf M},{\bf\bar M})_{N}$ & $\frac{3}{2}N$\\[1mm]
\hline
$+_{N,M,k}$ & $SU(N)\times SU(k)\times SU(M)^2 \times U(1)$   & $({\bf N},{\bf \bar k},{\bf 1},{\bf 1})_{M-\frac{N}{k}+1}$ & $\frac{3}{2}(M-\frac{N}{k} +1)$\\
&  & $({\bf 1},{\bf 1},{\bf M},{\bf \bar M})_N$ & $ \frac{3}{2}N$
\\[1mm]
\hline
$X_{N,M}$ & $SU(N)^2\,{\times}\, SU(M)^2\,{\times}\, U(1)$ & $({\bf N},{\bf \bar N},{\bf 1},{\bf 1})_{M}$ & $3M$\\
&&  $({\bf 1},{\bf 1},{\bf M},{\bf \bar M})_{N}$ & $3N$\\[1mm]
\hline
$\pslash_{N}$ & $SU(N)^6\,{\times}\, U(1)^3$ 
 & $({\bf N},{\bf\bar N},{\bf 1},{\bf 1},{\bf 1},{\bf 1})$ & $3N$\\
 && $({\bf 1},{\bf 1},{\bf N},{\bf\bar N},{\bf 1},{\bf 1})$  & $3N$\\
 &&  $({\bf 1},{\bf 1},{\bf 1},{\bf 1},{\bf N},{\bf\bar N})$ & $3N$\\
 &&  $({\bf N},{\bf 1},{\bf N},{\bf 1},{\bf N},{\bf 1})$ & $\frac{3}{2}(3N-1)$\\
 && $({\bf 1},{\bf N},{\bf 1},{\bf N},{\bf 1},{\bf N})$ & $\frac{3}{2}(3N-1)$\\
\bottomrule
\end{tabular}
\caption{The theories discussed in sec.~\ref{sec:5dscft}, with their global symmetries, stringy operators and the scaling dimensions inferred from field theory considerations. 
The proposed supergravity duals precisely reproduce the stringy operators and their scaling dimensions in the appropriate large-$N$ limits.\label{tab:CFT1}} 
\end{table}

The supergravity solutions corresponding to the brane webs are identified directly by their 5- and 7-brane charges. The geometry in the solutions of \cite{DHoker:2016ysh,DHoker:2017mds,DHoker:2017zwj} is a warped product of $AdS_6\times S^2$ over a Riemann surface $\Sigma$, and each solution is specified by a choice of two locally holomorphic functions $\cA_\pm$ on $\Sigma$.
Physically regular solutions were constructed for the case where $\Sigma$ is a disc, and the differentials $\partial\cA_\pm$ for these solutions have isolated poles on the boundary of $\Sigma$. These poles precisely represent the external 5-branes of the corresponding 5-brane web, with each pole representing a group of like-charged external 5-branes, where the residue encodes the charges $(p_i,q_i)$ and the number $N_i$ of external 5-branes. 
Though we cannot identify the full global symmetry of the 5d SCFT in the supergravity solution, we are able to identify the $U(1)$ factors in 
(\ref{eq:global-symmetry}). These correspond to reductions of the RR 4-form potential on 3-cycles surrounding the poles in the geometry, 
as will be explained in sec.~\ref{sec:3-cycles}.
For the solutions with additional 7-branes \cite{DHoker:2017zwj}, $\Sigma$ includes punctures around which the supergravity fields undergo non-trivial $SL(2,\RR)$ monodromy, with the monodromy representing the 7-brane charge and the position in $\Sigma$ representing the face of the web in which the 7-brane is placed.

Our main results, as described in section \ref{sec:strings-in-warped-AdS6}, will be to identify the states dual to the previously mentioned operators in terms of strings or string-webs embedded into these solutions.
With $AdS_6$ in global coordinates, such that the SCFTs are realized on the cylinder $\RR\times S^4$, the  strings and string webs are localized at the origin 
in the radial coordinate of $AdS_6$, and extend along a one-dimensional subspace of $\Sigma$, that connects poles or poles and punctures. 
The scaling dimensions of the dual operators can be obtained directly from the Hamiltonian. We also explicitly obtain their $R$-symmetry charge, from the coupling to the bulk gauge field dual to the $R$-symmetry current in the SCFTs, for which we identify the relevant part in the corresponding supergravity fluctuation. 
Their global $U(1)$ charges are easily read off from the poles or punctures on $\Sigma$ that the strings end on. 
This information shows that the string states are BPS saturated, and agree with the field theory results on the scaling dimensions in the large-$N$ limit. For the $+_{N,M}$ theory, we also discuss explicitly how a chiral ring relation is recovered in the supergravity dual.

In summary, we have a precise quantitative match between field theory analyses and holographic computations of the spectrum of a class of large-scaling-dimension operators.
The results support the identifcation of the Type IIB supergravity solutions of \cite{DHoker:2017mds,DHoker:2017zwj} with the proposed 5d SCFTs. For the future, it would be desirable to further substantiate this match and to further exploit the holographic descriptions to better understand this class of 5d SCFTs. For example, with a compelling case for the precise form of the dual SCFTs, the holographic computations of the $S^5$ partition functions in \cite{Gutperle:2017tjo,Gutperle:2018vdd} may be seen as predictions for the dual SCFTs, and it would be an interesting further check to test them.

\subsection{Outline}
The rest of the paper is organized as follows. In sec.~\ref{sec:5dscft} we present a number of case studies of 5d SCFTs described by simple 5-brane webs,
and describe the spectrum of stringy operators in each case.
In sec.~\ref{sec:sugra-review} we give a brief review of the warped $AdS_6\times S^2\times\Sigma$ Type IIB supergravity solutions, discuss the charge quantization and identify, in parts, a fluctuation dual to the conserved SCFT $R$-symmetry current. We also identify the bulk gauge fields dual to the $U(1)$ factors in (\ref{eq:global-symmetry}). In sec.~\ref{sec:strings-in-warped-AdS6} we discuss $(p,q)$ strings and string webs embedded into the supergravity solutions dual to the SCFTs of sec.~\ref{sec:5dscft}, and identify them with the operators discussed there.

\section{5d SCFT case studies}\label{sec:5dscft}

In this section we consider a number of examples of 5d SCFT's that admit mass deformations
leading to 5d supersymmetric quiver gauge theories with $SU$ gauge symmetries.

\subsection{The \texorpdfstring{$T_N$}{T-N} theory} \label{sec:T-N}

\begin{figure}
\center
\includegraphics[height=0.3\textwidth]{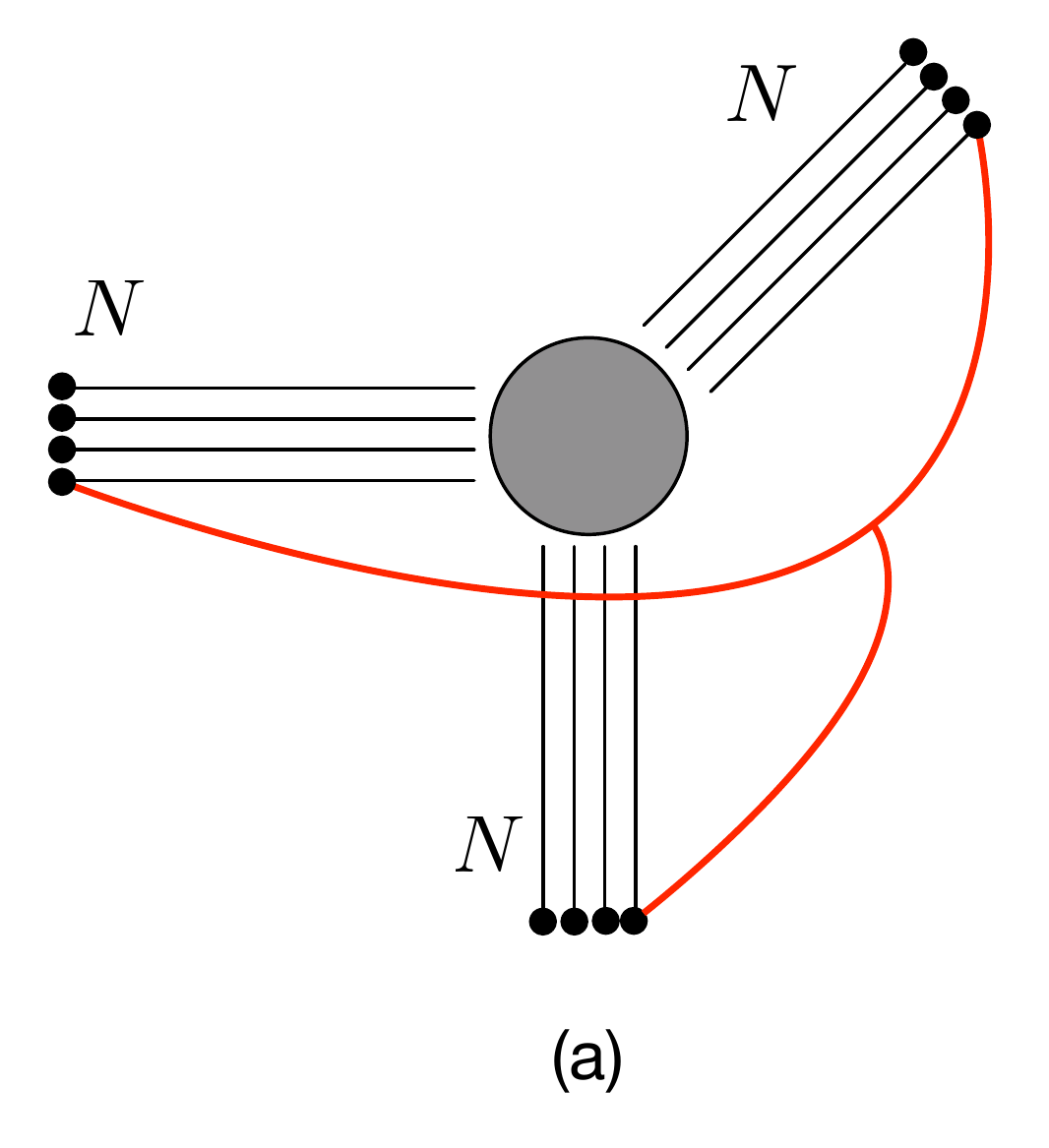}
\hspace{10pt}
\includegraphics[height=0.3\textwidth]{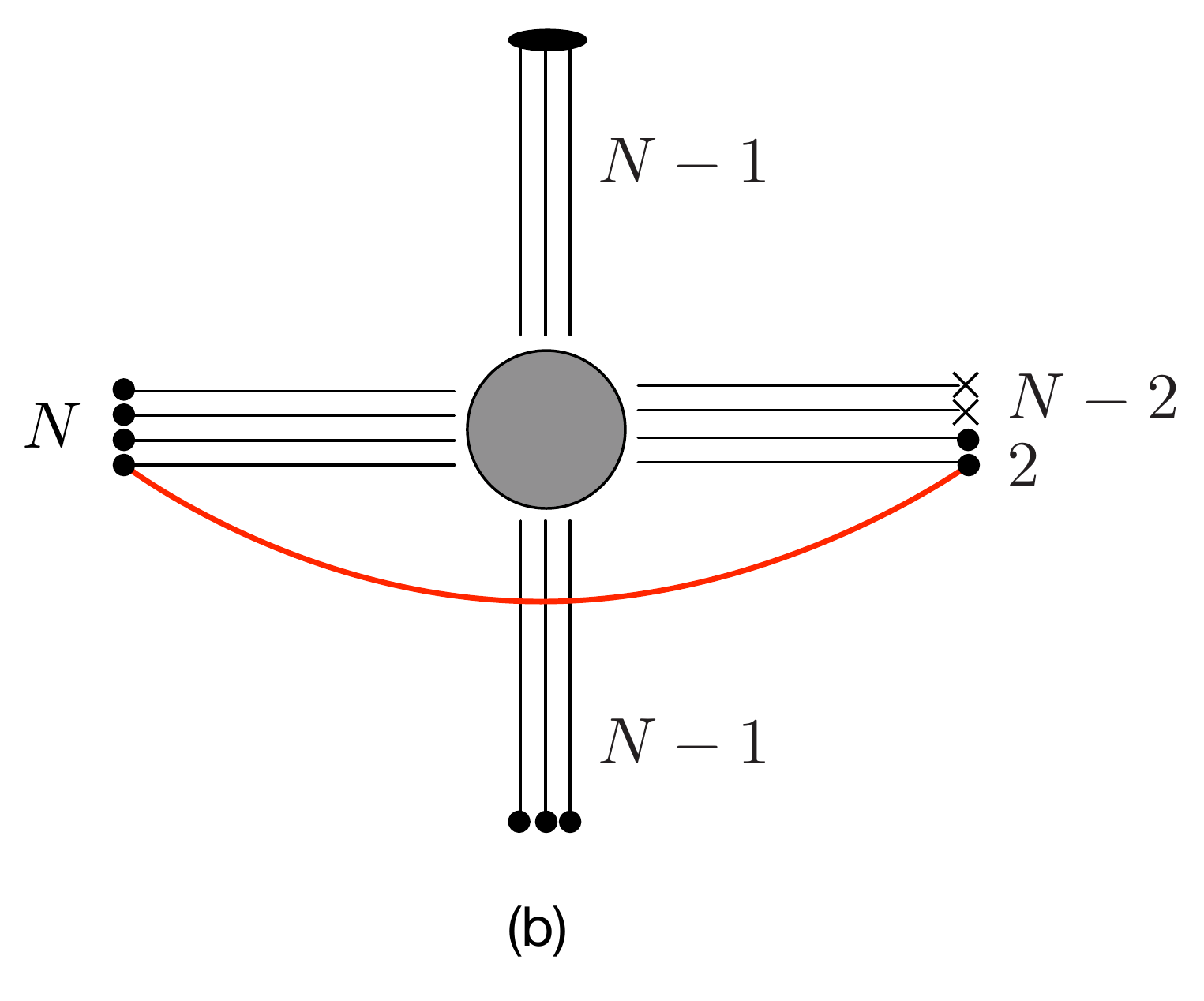}
\hspace{10pt}
\includegraphics[height=0.35\textwidth]{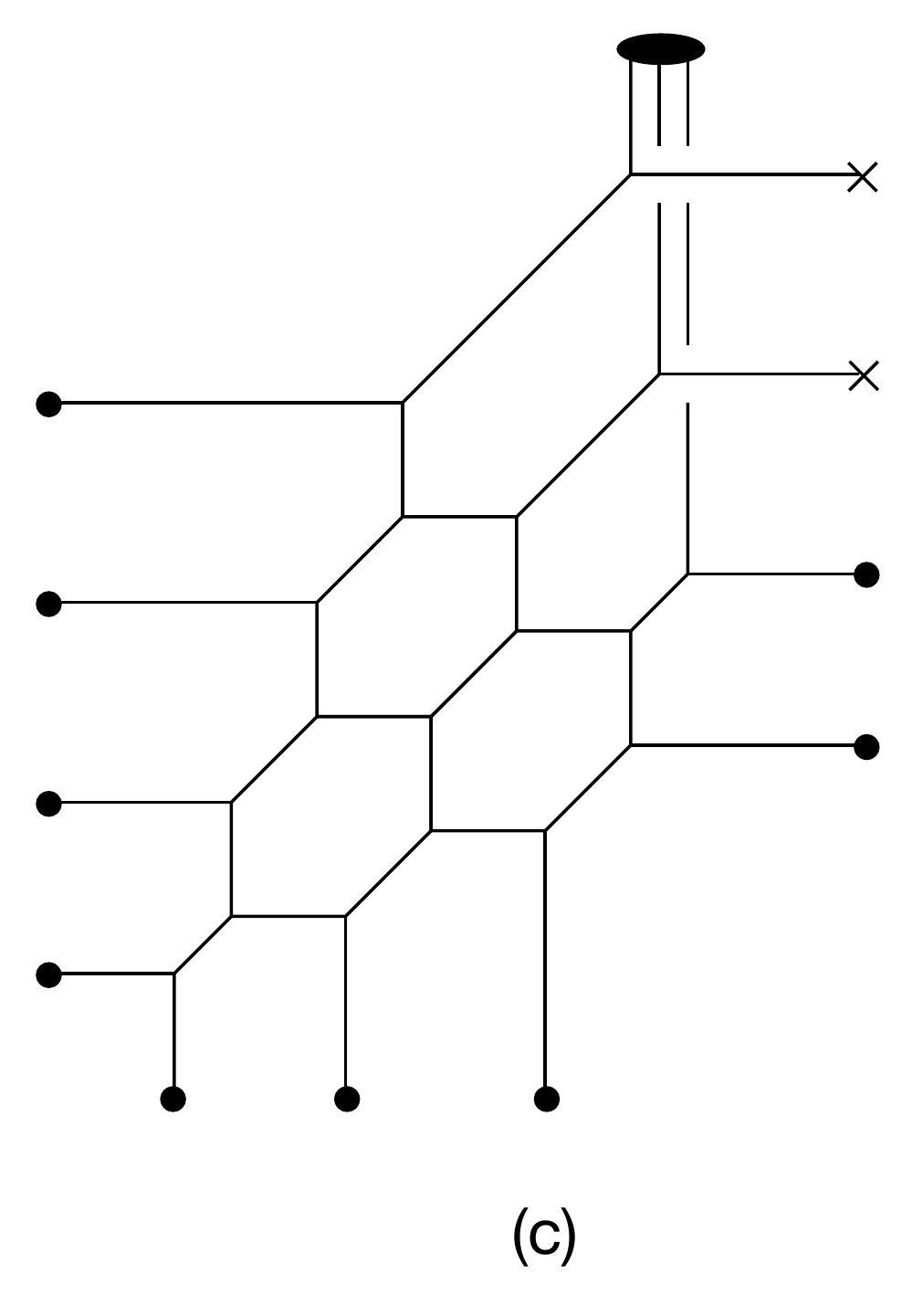} 
\caption{Brane webs for the 5d $T_N$ theory, with the string junction/open strings representing the long operators. The non-flavor D7-branes are denoted by an ``x" rather than by a dot. \label{fig:TN}}
\label{TNjunction}
\end{figure}

The 5d $T_N$ theory corresponds to a triple junction of $N$ D5-branes, $N$ NS5-branes,  and $N$ (1,1) 5-branes, shown in fig.~\ref{fig:TN}(a). 
The global symmetry of this theory is in general $SU(N)^3$.
For $N=2$ this is just the theory of four free hypermultiplets, which has a global symmetry $Sp(4)$,
and for $N=3$ the global symmetry is enhanced to $E_6$.
This theory reduces to the 4d $T_N$ theory upon compactification on $S^1$ \cite{Benini:2009gi}.
The spectrum of chiral operators of the 4d $T_N$ theory contains a scalar operator
in the tri-fundamental $({\bf N},{\bf N},{\bf N})$ representation of $SU(N)^3$ with a scaling dimension
$\Delta = N-1$ \cite{Gaiotto:2009gz}.\footnote{More generally there are operators in the 
tri-$k$-antisymmetric representation 
$({\bf N}^k_{asym},{\bf N}^k_{asym},{\bf N}^k_{asym})$ with $k=1,\ldots, N-1$ and scaling
dimension $\Delta=\frac{3}{2}\,k\,(N-k)$ \cite{Hayashi:2014hfa}.}
The corresponding operator in the five dimensional theory has a scaling dimension $\Delta = \frac{3}{2}(N-1)$.
In particular for $N=2$ this is a free field corresponding to the four free hypermultiplets of the $T_2$ theory,
and for $N=3$ it corresponds to a conserved current multiplet in the $({\bf 3},{\bf 3},{\bf 3})$ representation of $SU(3)^3$, giving the $E_6$
global symmetry of the $T_3$ theory.
The $({\bf N},{\bf N},{\bf N})$ operator
is naturally described in the 5-brane web construction as a 3-pronged-string connecting a $(1,0)$ 7-brane,
a $(0,1)$ 7-brane, and a $(1,1)$ 7-brane, as shown in fig.~\ref{TNjunction}a.

The $T_N$ theory admits a relevant deformation which flows to the quiver gauge theory given by (see fig.~\ref{TNjunction}) 
\cite{Bergman:2014kza,Hayashi:2014hfa}
\be
\label{TNquiver}
[2]\stackrel{x_1}{-}(2)\stackrel{x_2}{-}(3)-\cdots-(N-2)\stackrel{x_{N-2}}{-}(N-1)\stackrel{x_{N-1}}{-}[N] \,,
\ee
where each element $(k)$ corresponds to an $SU(k)$ gauge symmetry, and each element $[k]$ to an $SU(k)$ global symmetry.
We will use $a,b$ to denote global indices and $\alpha,\beta$ to denote gauge indices.
We also use $\lbrace x_i,\tilde x_i\rbrace$ to denote the scalars in the matter hypermultiplet corresponding to the $i$-th link.
The global symmetry of the gauge theory is $SU(N)\times SO(4)\times U(1)_B^{N-2}\times U(1)_I^{N-2}$,
where the $U(1)_B$'s are associated to the matter fields, and the $U(1)_I$'s are the topological symmetries associated to the 
simple gauge group factors.
At the fixed point the global symmetry is enhanced by instantons to $SU(N)^3$.
This was shown for $N=4,5$ in \cite{Bergman:2014kza}.
For $N=3$ the symmetry is further enhanced to $E_6$, as shown in \cite{Kim:2012gu}.

Some of the components of the tri-fundamental operator can be described in terms of the matter fields of the IR quiver gauge theory.
We can form several gauge invariant dimension $\frac{3}{2}(N-1)$ operators as follows:
\be
{\cal O}^a_{\tilde{b}} & = &  [x_{1} \cdots x_{N-1}]^a_{\tilde{b}} \\
{\cal O}_{a\tilde{b}} & = & \epsilon_{\alpha\beta} \, [\tilde{x}_1]^\alpha_a \, [x_{2} \cdots x_{N-1}]^\beta_{\tilde{b}} \\
{\cal O}_{(j)\tilde{b}} &=& [\det \tilde{x}_j]_\alpha \, [x_{j+1} \cdots x_{N-1}]^\alpha_{\tilde{b}} \\
{\cal O}_{(N-1)\tilde{b}} &=& [\det \tilde{x}_{N-1}]_{\tilde{b}}
\ee
where $a$ is a flavor $SU(2)$ index, $\tilde{b}$ is a flavor $SU(N)$ index, $j = 2,\ldots, N-2$ and
we use the shorthand 
\be 
[\det \tilde{x}_j]_\alpha \equiv \epsilon_{\alpha\alpha_1\cdots\alpha_j} \epsilon^{\beta_1\cdots \beta_j} \,
\tilde{x}_{\beta_1}^{\alpha_1} \cdots \tilde{x}_{\beta_j}^{\alpha_j} \,,
\ee
and likewise for $[\det \tilde{x}_{N-1}]_{\tilde{b}}$.
The determinant of the bi-fundamental field has a left-over index since neighboring groups differ by one.
All of these operators transform in the ${\bf N}$ representation of the global $SU(N)$ symmetry of the IR gauge theory,
and correspond to a subset of the components of the $SU(N)^3$ tri-fundamental operator at the fixed point.
In the 5-brane web of fig.~\ref{TNjunction}b they simply correspond to an open string between a D7-brane and on the right and a D7-brane on the left,
which can be traced back to the 3-pronged string in the original 5-brane web via a Hanany-Witten transition.
The remaining components of the tri-fundamental operator will naturally involve instanton operators.

\subsection{The \texorpdfstring{$Y_N$}{Y-N} theory}\label{sec:Y-N}

\begin{figure}
\center
\includegraphics[height=0.25\textwidth]{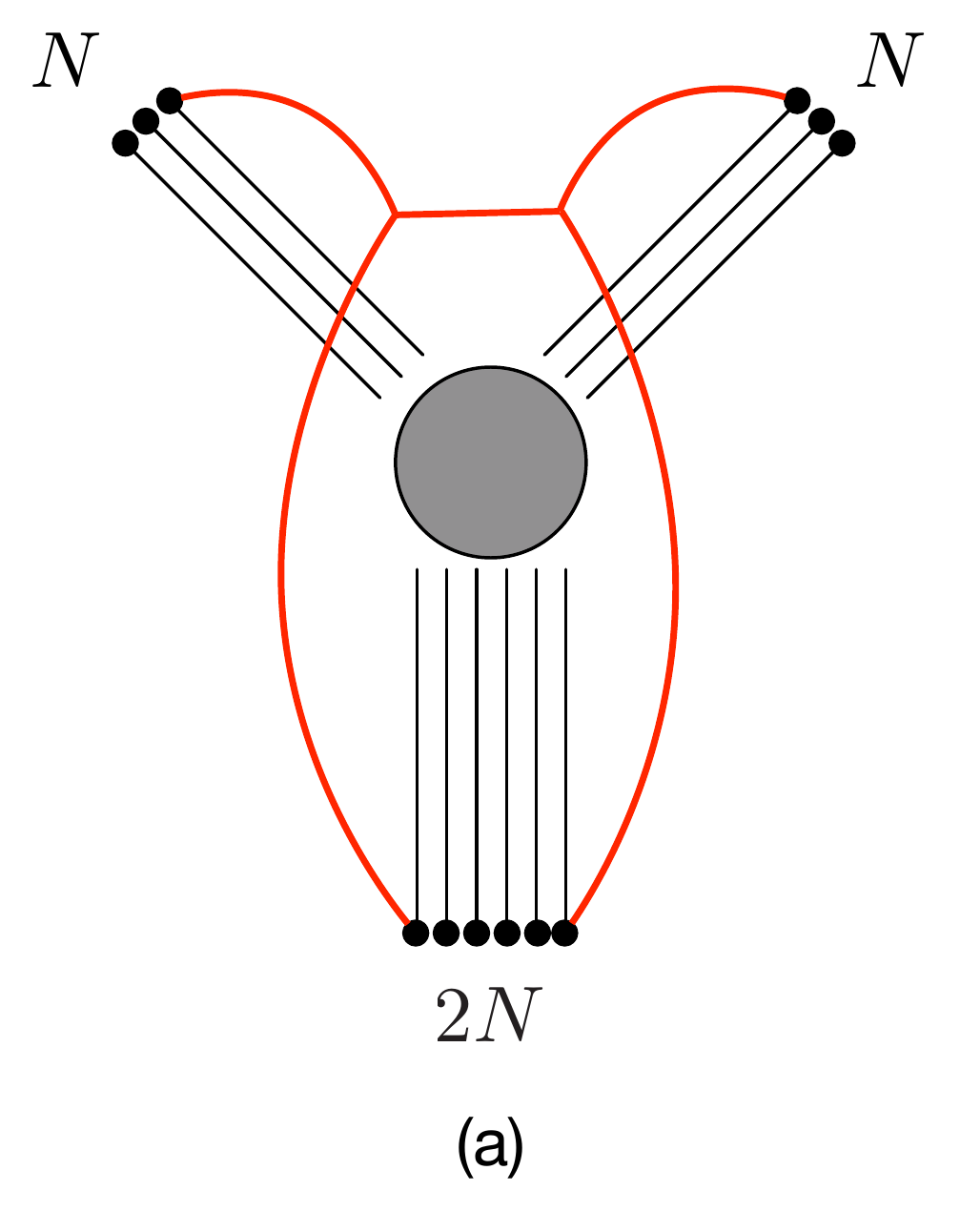}
\hspace{10pt}
\includegraphics[height=0.28\textwidth]{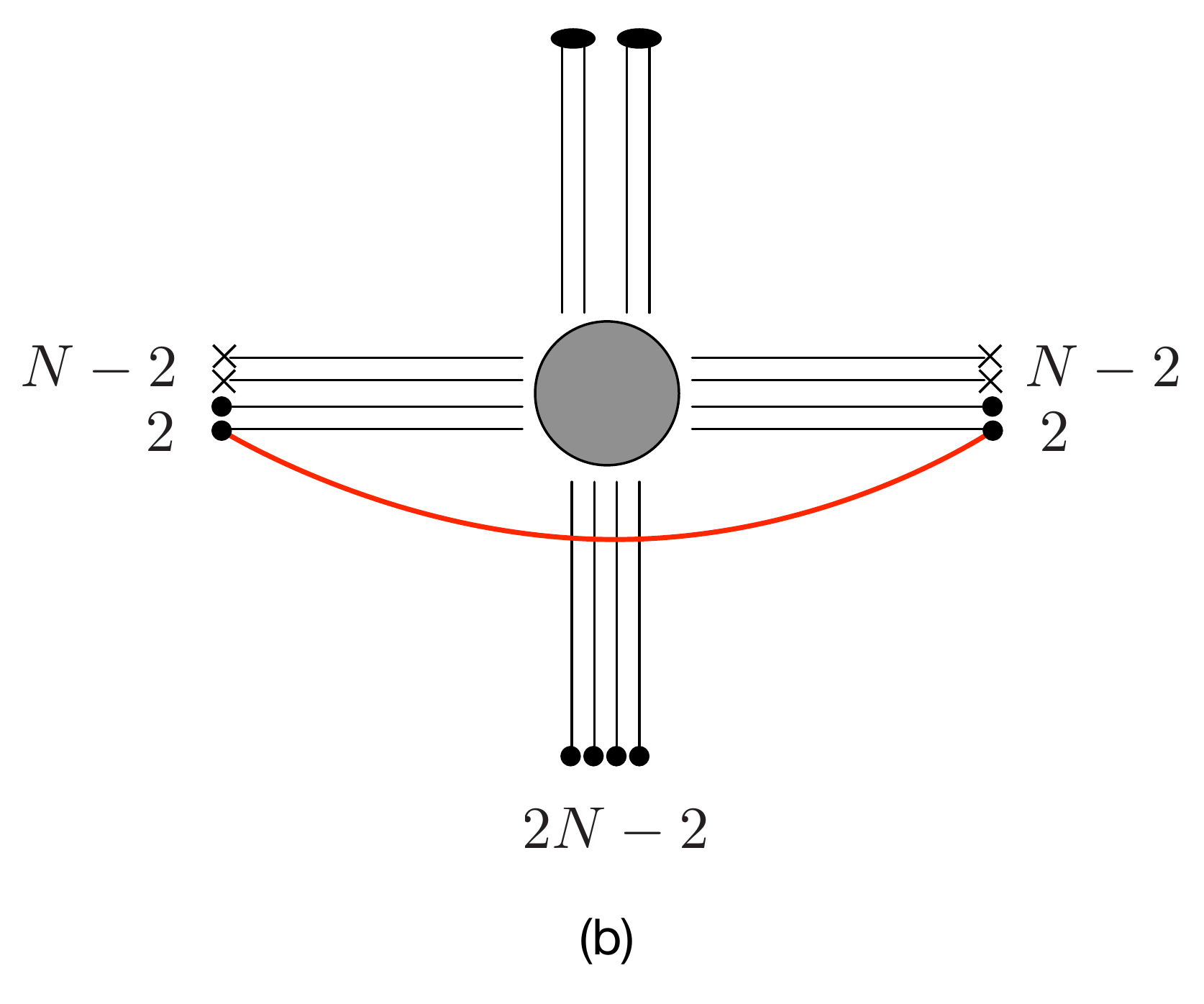} 

\caption{Brane webs for the $Y_N$ theory with $N=3$, with the string junction/open strings representing the long operators.}
\label{YNjunction}
\end{figure}

Having gained confidence in the identification of the stringy operators of the $T_N$ theory, we move on to less
familiar ground, and consider a number of other theories.
We begin with another triple junction, this time of $2N$ $(0,1)$ (NS) 5-branes, $N$ $(1,1)$ 5-branes, and $N$ $(-1,1)$ 5-branes, 
as shown in fig.~\ref{YNjunction}(a).
The global symmetry of the so-called $Y_N$ theory is in general $SU(2N)\times SU(N)^2$.
For $N=2$ this is actually the $E_5$ theory, in which the global symmetry is enhanced to $E_5 = SO(10)$.
We claim that the spectrum of chiral operators of the $Y_N$ theory contains a scalar operator
in the $(({\bf 2N})^2_{asym},{\bf N},{\bf N})$ representation of the global symmetry.
This operator is represented by a string-web consisting of two external $(0,1)$ strings, an external $(1,1)$ string, and an external $(1,-1)$ string,
connected by an internal $(1,0)$ string, as shown in fig.~\ref{YNjunction}(a).
The fundamental charges under the two $SU(N)$ factors are obvious.
The reason for the antisymmetric representation under $SU(2N)$ is that
the two $(0,1)$ strings are constrained by the {\em s-rule} to attach to two different $(0,1)$ 7-branes.
Unlike in the $T_N$ theory, the dimension of this operator in the $Y_N$ theory is not known a-priori.
However, by looking at the IR quiver gauge theories we will see that $\Delta = 3(N-1)$.
This is also consistent with the $N=2$ case, where this operator has $\Delta = 3$ and transforms in the $({\bf 6},{\bf 2},{\bf 2})$ representation
of $SU(4)\times SU(2)^2$, providing the additional conserved current 
multiplet responsible for the enhancement of $SU(4)\times SU(2)^2 = SO(6)\times SO(4)$ to $E_5 = SO(10)$.

One of the IR quiver gauge theories can be obtained by separating the NS5-branes, and
is most easily read-off from an equivalent web obtained by moving two of the $(0,1)$ 7-branes upwards, fig.~\ref{YNjunction}(c).
The resulting gauge theory is given by
\be
[2]\stackrel{x_1}{-}(2)\stackrel{x_2}{-}(3)-\cdots-(N-1)\stackrel{x_{N-1}}{-}(N)\stackrel{x_N}{-}(N-1)-\cdots-(3)\stackrel{x_{2N-3}}{-}(2)\stackrel{x_{2N-2}}{-}[2] \,. \nonumber \\
\ee
The global symmetry of this theory is $SO(4)\times SO(4) \times U(1)_B^{2N-4} \times U(1)_I^{2N-3}$. 
This should enhance to $SU(2N)\times SU(N)^2$ at the fixed point.
Note that for $N=2$ the IR theory reduces to $SU(2)$ with four flavors, which is the IR theory corresponding to the $E_5$ fixed point.

The string-web connected to the two 7-branes that were moved becomes an open string connecting a D7-brane on the left
to a D7-brane on the right.
This describes three types of operators in the IR quiver gauge theory, all of scaling dimension $\Delta = 3(N-1)$.
The first type involves both end nodes, and includes
\be
{\cal O}^a_{\tilde{b}} &=& [x_1 \cdots x_{2N-2}]^a_{\tilde{b}} \\
{\cal O}^{a\tilde{b}} & = & \epsilon^{\alpha\beta} \, [x_{1} \cdots x_{2N-3}]^a_\alpha \, (\tilde{x}_{2N-2})^{\tilde{b}}_\beta \\
{\cal O}_{a\tilde{b}} &=& \epsilon_{\alpha\beta} \, (\tilde{x}_1)^\alpha_a \, [x_2 \cdots x_{2N-2}]^\beta_{\tilde{b}} \\
{\cal O}_{a}^{\tilde{b}} &=& \epsilon_{\alpha\beta} \, \epsilon^{\gamma\delta} \, (\tilde{x}_1)^\alpha_a \, [x_2 \cdots x_{2N-3}]^\beta_\gamma \,
(\tilde{x}_{2N-2})^{\tilde{b}}_\delta
\ee
where $a$ and $\tilde{b}$ are indices of the two flavor $SU(2)$'s, respectively.
Together these transform in the $({\bf 4},{\bf 4})$ representation of $SO(4)\times SO(4)$.
The second type involves a single end node, and includes
\be
{\cal O}^{a(j)} &=&  [x_{1}\cdots x_{2N-j-1}]^a_\alpha \, [\det \tilde{x}_{2N-j}]^\alpha \\
{\cal O}_a^{(j)} &=& \epsilon_{\alpha\beta} \, (\tilde{x}_1)^\alpha_a \, [x_{2}\cdots x_{2N-j-1}]^\beta_\gamma \, [\det \tilde{x}_{2N-j}]^\gamma \\
{\cal O}_{\tilde{b}}^{(j)} &=&  [\det \tilde{x}_{j-1}]_\alpha \, [x_{j}\cdots x_{2N-2}]^\alpha_{\tilde{b}}\\
{\cal O}^{\tilde{b}(j)} &=& \epsilon^{\alpha\beta} \, [\det \tilde{x}_{j-1}]_\gamma \, [x_{j}\cdots x_{2N-3}]^\gamma_\alpha \,
(\tilde{x}_{2N-2})^{\tilde{b}}_\beta
\ee
where $j=3, \ldots, N$.
These give $N-2$ operators in the $({\bf 4},{\bf 1})$ representation and $N-2$ in the $({\bf 1},{\bf 4})$ representation.
The third type does not involve the end nodes, and is given by 
\be
{\cal O}^{(ij)} = [\det \tilde{x}_i]_\alpha \, [x_{i+1}\cdots x_{N+j-3}]^\alpha_\beta \, [\det \tilde{x}_{N+j-2}]^\beta \,,
\ee
where $i,j=3,\ldots , N$.
These are $(N-2)^2$ operators in the $({\bf 1},{\bf 1})$ representation.
Altogether these operators provide $(N+2)^2$ of the components of the $(({\bf 2N})^2_{asym},{\bf N},{\bf N})$ operator.
The rest will require the inclusion of instantons.

\begin{figure}
\center
\includegraphics[height=0.24\textwidth]{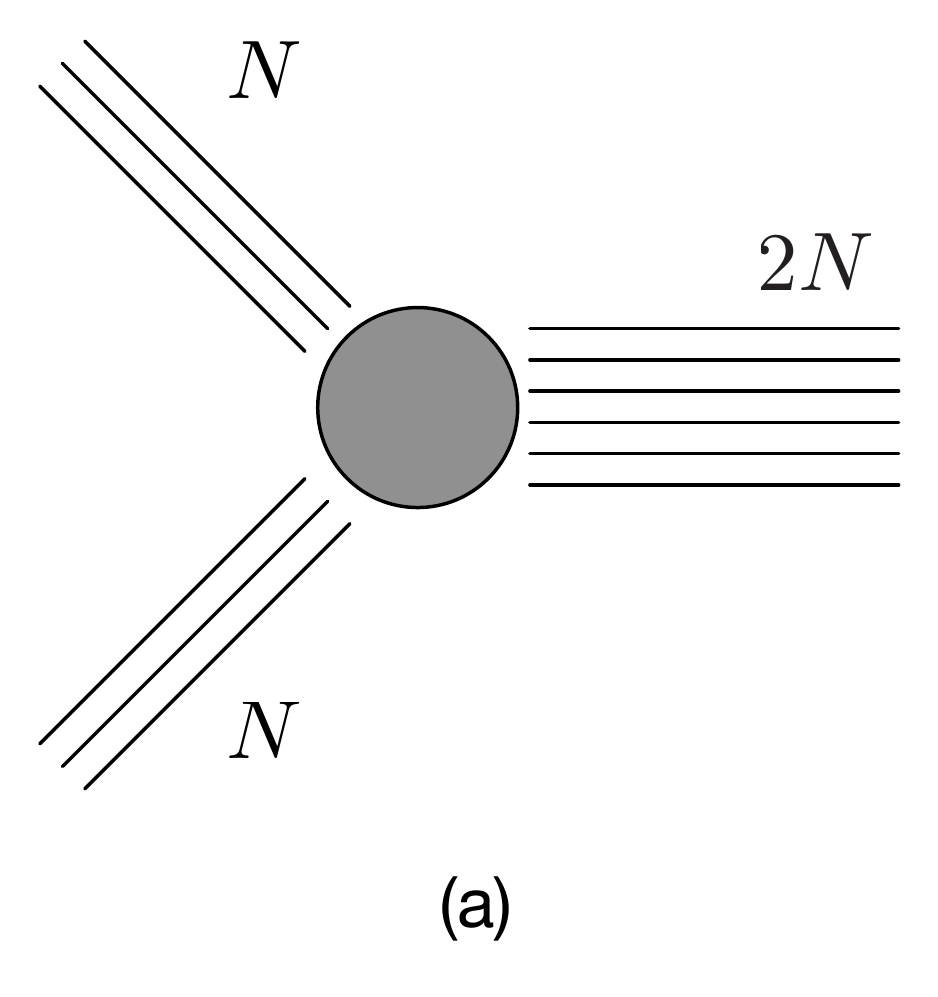} 
\hspace{40pt}
\includegraphics[height=0.24\textwidth]{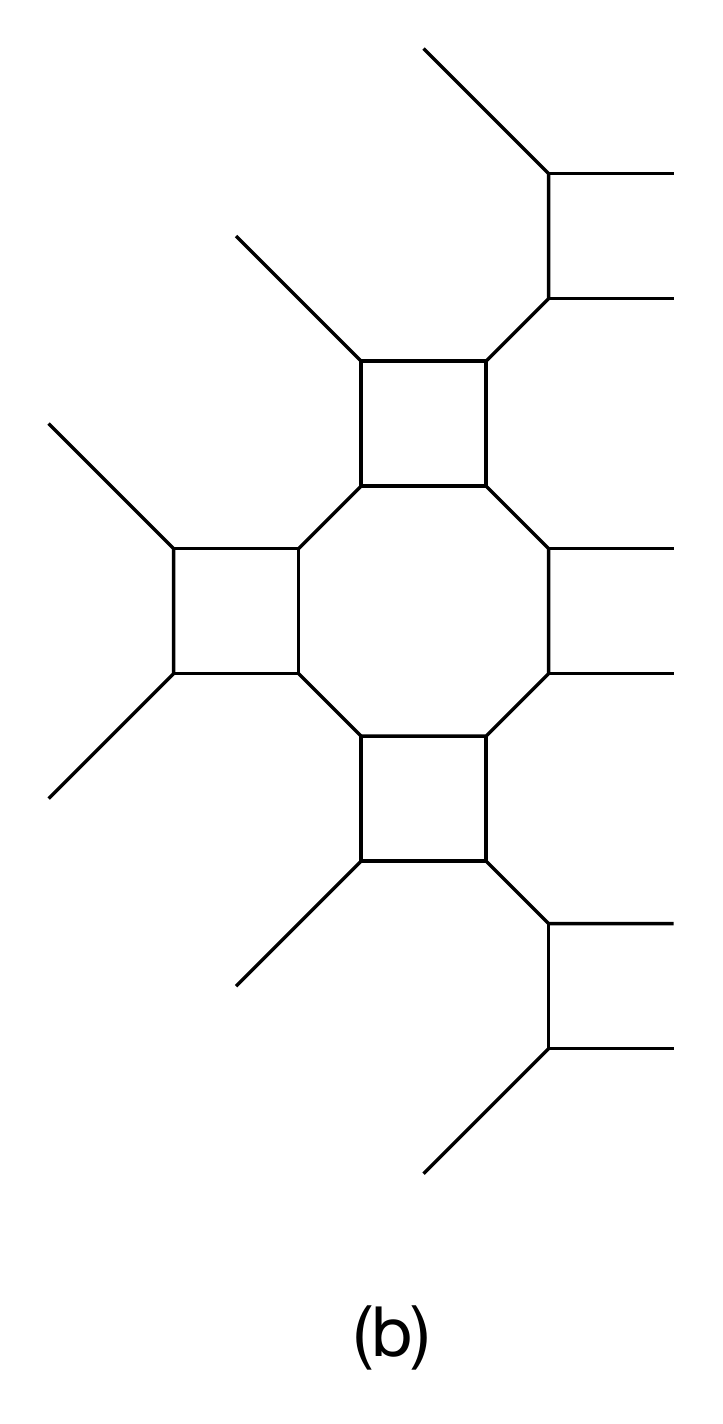} 
\caption{The S-dual of the $Y_N$ web shown in fig.~\ref{YNjunction}, with a deformation to a quiver gauge theory.}
\label{DualYNjunction}
\end{figure}

It is also interesting to consider a second IR quiver gauge theory, corresponding to the s-dual web,  (fig.~\ref{DualYNjunction}).
In this case the gauge theory is given by
\be
\label{YNQuiver2}
(2)\stackrel{y_1}{-}(4)\stackrel{y_2}{-}\cdots\stackrel{y_{N-2}}{-}(2N-2)\stackrel{y_{N-1}}{-}[2N] \,.
\ee
In this description the $SU(2N)$ global symmetry is manifest, and we can explicitly construct $N$ operators of
dimension $\Delta = 3(N-1)$ in the $({\bf 2N})^2_{asym}$ representation of $SU(2N)$ as follows:
\be
{\cal O}_{ab(1)} &=& \epsilon_{\alpha\beta} \, [y_1 \cdots y_{N-1}]_a^\alpha \, [y_1 \cdots y_{N-1}]_b^\beta \\
{\cal O}_{ab(j)} &=& [\det \tilde{y}_j]_{\alpha\beta} \,
 [y_{j+1} \cdots y_{N-1}]_a^\alpha \, [y_{j+1} \cdots y_{N-1}]_b^\beta  \\ 
 {\cal O}_{ab(N-1)} &=& [\det \tilde{y}_{N-1}]_{ab} 
\ee
where $j=1, \ldots, N-2$. Here the determinant of the bi-fundamental has two left-over indices since neighboring groups differ by two.

\subsection{The \texorpdfstring{$+_{N,M}$}{+-N-M} theory} \label{sec:grid-N-M}

\begin{figure}
\center
\subfigure[][]{\label{fig:grid-1}
\includegraphics[height=0.29\textwidth,trim={0cm 1.35cm 0cm 0cm},clip]{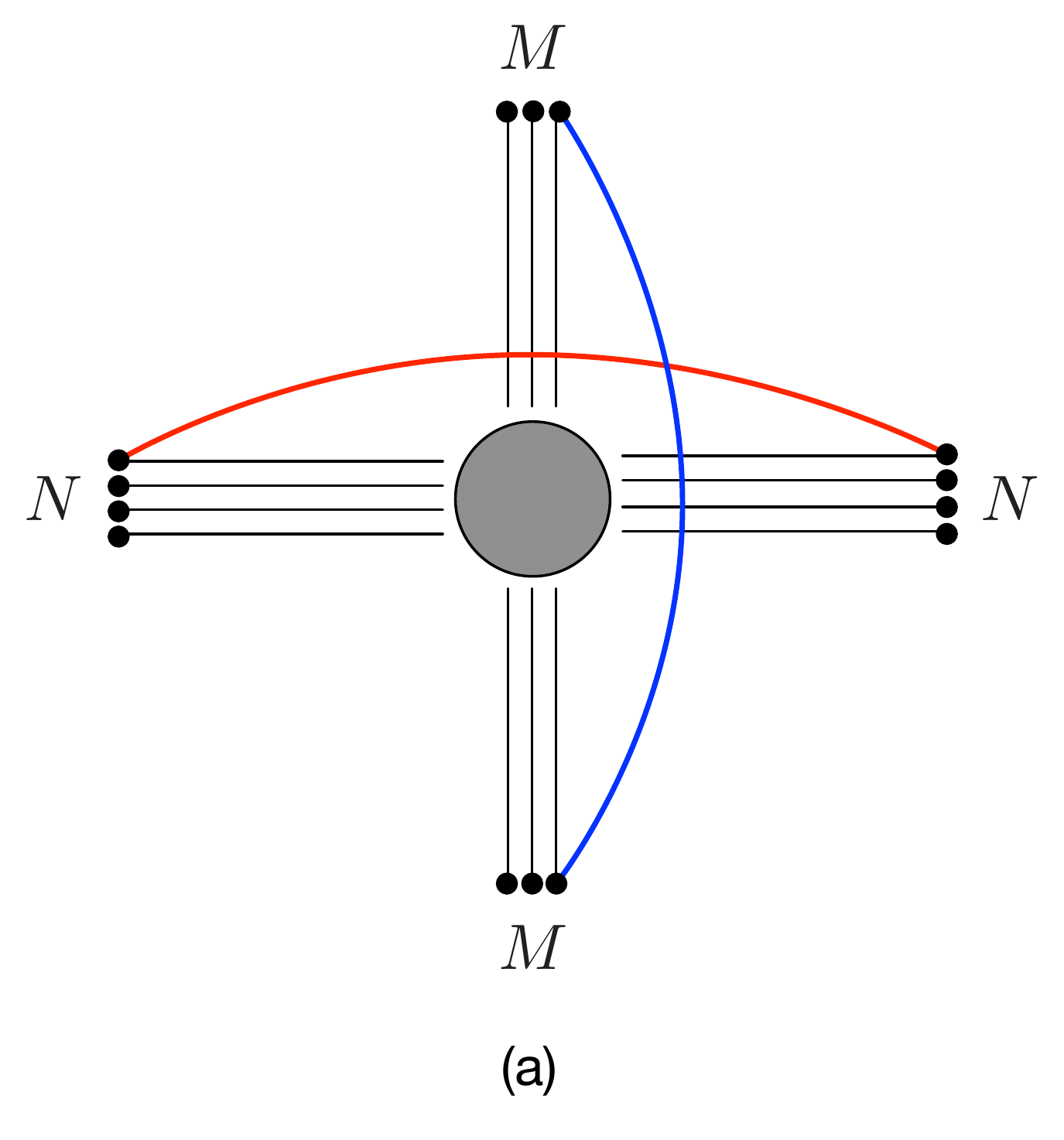} 
}\hspace{8pt}
\subfigure[][]{\label{fig:grid-2}
 \includegraphics[height=0.29\textwidth,trim={0cm 1.35cm 0cm 0cm},clip]{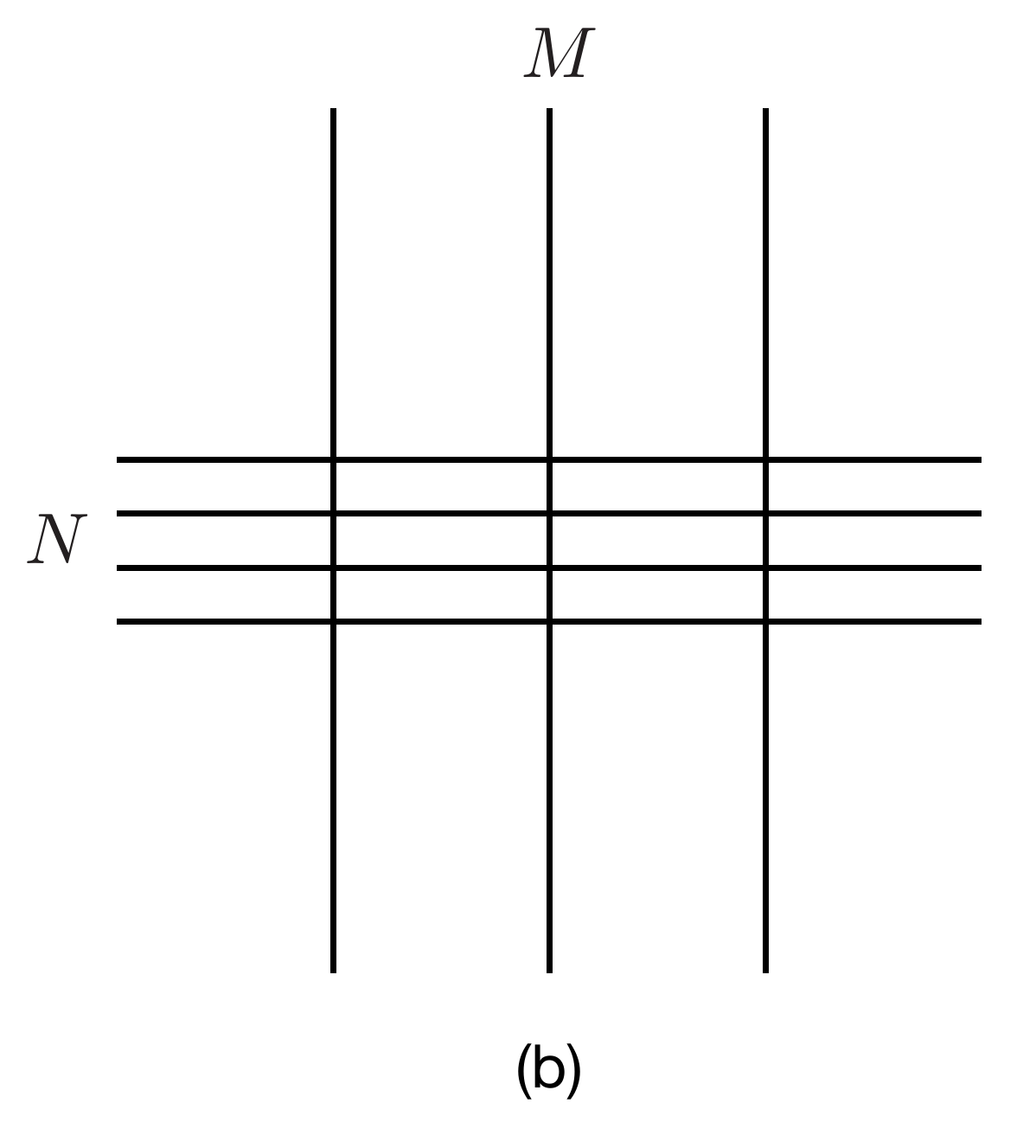} 
}\hspace{8pt}
\subfigure[][]{\label{fig:grid-3}
\includegraphics[height=0.29\textwidth,trim={0cm 1.35cm 0cm 0cm},clip]{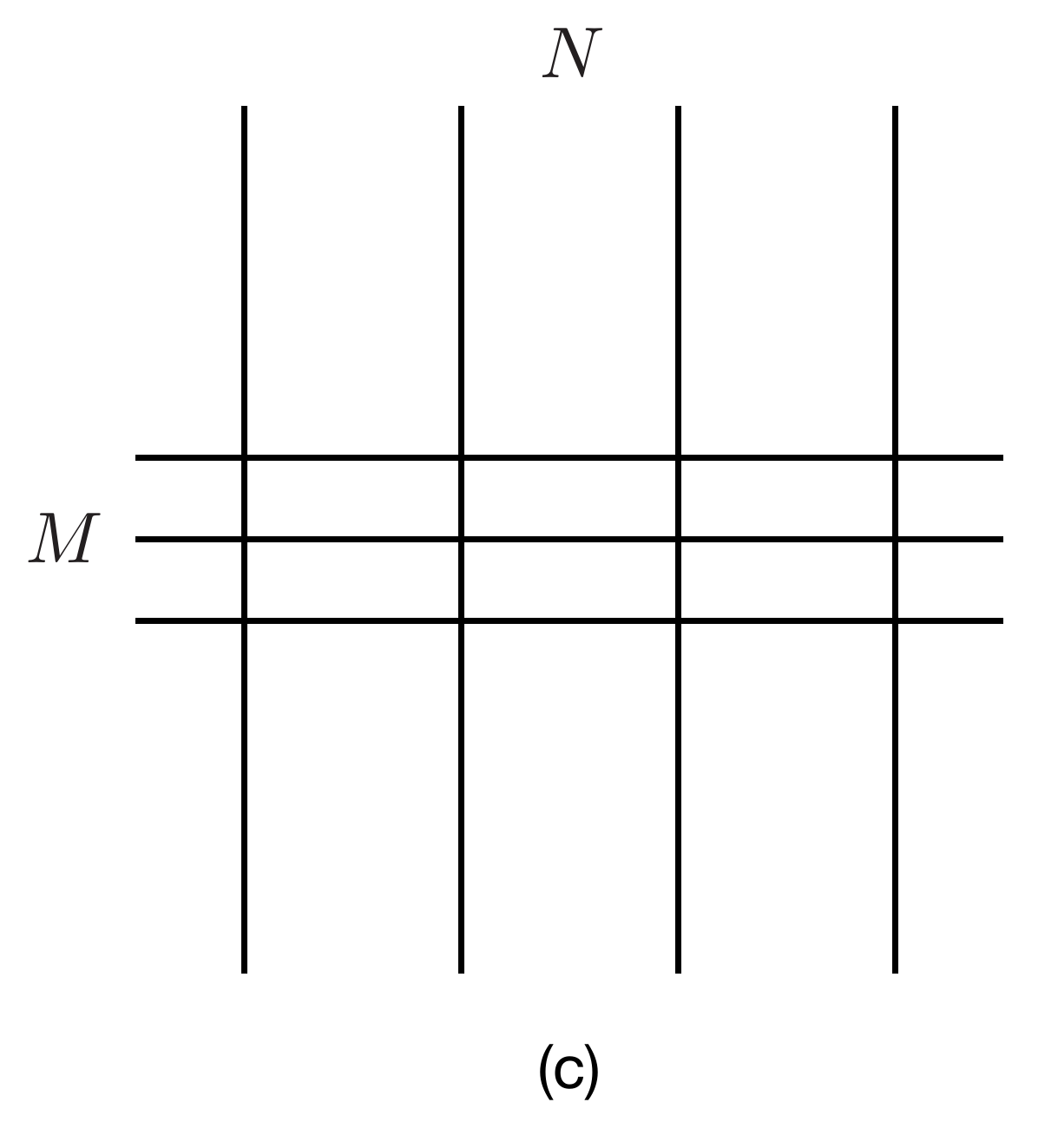}
}
\caption{The 5-brane web of the $+_{N,M}$ theory: (a) The theory at the fixed point, with the string and D1-brane representing the bi-fundamental operators. 
(b) The deformation leading to the $SU(N)$ quiver.
(c) The deformation leading to the $SU(M)$ quiver.}
\end{figure}

Next consider the quartic junction of two sets of $N$ D5-branes and two sets of $M$ NS5-branes, fig.~\ref{fig:grid-1},
which we call the $+_{N,M}$ theory.
Some aspects of this theory were originally studied in \cite{Aharony:1997bh}.
The 5-brane web construction suggests that the global symmetry is $SU(N)^2\times SU(M)^2 \times U(1)$,
and that there are chiral operators transforming in the bi-fundamental representations
$({\bf N},\bar{\bf N},{\bf 1},{\bf 1})_{M}$ and $({\bf 1},{\bf 1},{\bf M},\bar{\bf M})_{N}$ and their conjugates, corresponding, respectively, 
to open strings between (1,0) 7-branes, and D1-branes between (0,1) 7-branes.
By identifying subsets of these operators in the IR quiver gauge theories we will see that their scaling dimensions
are given by $\Delta_{({\bf N},\bar{\bf N})} = \frac{3}{2}M$ and $\Delta_{({\bf M},\bar{\bf M})} = \frac{3}{2}N$.
This is consistent in particular with the special case of $M=1$ which corresponds to $N^2$ free hypermutiplets, all
of dimension $\Delta_{({\bf N},\bar{\bf N})} =\frac{3}{2}$.
It is also consistent with the case of $M=2$, where the $SU(N)^2\times U(1)$ part of the global symmetry is enhanced to 
$SU(2N)$ \cite{Aharony:1997bh}.
In this case the $({\bf N},\bar{\bf N})$ operator has a scaling dimension $\Delta_{({\bf N},\bar{\bf N})} = 3$, and provides
the extra conserved currents responsible for the enhancement.
If both $N=M=2$, the $+_{N,M}$ theory is the $E_5$ theory, in which the global symmetry is further enhanced to $E_5 = SO(10)$.
In this case both operators have a scaling dimension $\Delta = 3$, and transform in the 
$({\bf 2},{\bf 2},{\bf 1},{\bf 1})_{\pm 2} + ({\bf 1},{\bf 1},{\bf 2},{\bf 2})_{\pm 2}$ of $SU(2)^4\times U(1)$.
These are some but not all of the conserved currents required for the enhancement.
An operator with charges $({\bf 2},{\bf 2},{\bf 2},{\bf 2})_0$ is missing.
This motivates us to conjecture the existence of a chiral operator in the $({\bf N},\bar{\bf N},{\bf M},\bar{\bf M})_0$
representation with scaling dimension $\Delta = \frac{3}{4}NM$.
However this does not correspond to strings in the 5-brane web.

The IR quiver gauge theory resulting from separating the NS5-branes is given by
\be
\label{GridQuiver1}
[N]\stackrel{x_1}{-}(N)\stackrel{x_2}{-}\cdots-(N)\stackrel{x_M}{-}[N] \,,
\ee
with the $SU(N)$ gauge node appearing $M-1$ times.
This has in general a global symmetry $SU(N)^2 \times U(1)_B^{M} \times U(1)_I^{M-1}$.
For $M=2$ the global symmetry is enhanced to $SU(2N)\times U(1)_B \times U(1)_I$.
The dual quiver gauge theory resulting from separating the D5-branes is given by
\be
\label{GridQuiver2}
[M]\stackrel{y_1}{-}(M)\stackrel{y_2}{-}\cdots-(M)\stackrel{y_N}{-}[M] \,,
\ee
with the $SU(M)$ gauge node appearing $N-1$ times.
This has in general a global symmetry $SU(M)^2 \times U(1)_B^{N} \times U(1)_I^{N-1}$,
which is enhanced for $N=2$ to $SU(2M)\times U(1)_B \times U(1)_I$.
In both cases the global symmetry is expected to enhance at the UV fixed point to $SU(N)^2\times SU(M)^2\times U(1)$,
except when either $N = 2$ or $M=2$ or both.
This was demonstrated for $M=3$ in \cite{Tachikawa:2015mha} by counting 1-instanton states.
The remaining $U(1)$ factor is the sum of all the $U(1)_B$ symmetries.

Some components of the bi-fundamental operators are realized in the IR quivers as follows.
In the first quiver theory (\ref{GridQuiver1}) we have a dimension $\frac{3}{2}M$ 
operator
\be
{\cal O}^a_{\tilde{b}} = [x_1 \cdots x_{M}]^a_{\tilde{b}}\,,
\ee
transforming in the $({\bf N},\bar{\bf N})$ representation of $SU(N)^2$,
and dimension $\frac{3}{2}N$ singlets
\be
{\cal O}_{(i)} = \det x_i \,,
\ee
with $i = 1, \ldots, M$.
In addition, the former carries $M$ units of charge under the sum of the $U(1)_B$ symmetries, and the latter carries $N$ units of charge.
These operators satisfy a ``chiral ring relation":
\be\label{eq:chiral-ring}
\prod_{i=1}^M {\cal O}_{(i)} = \det {\cal O}^a_{\tilde{b}} \,.
\ee
Likewise, the second quiver theory (\ref{GridQuiver1}) has a dimension $\frac{3}{2}N$
operator $\tilde{\cal O}^a_{\tilde{b}}$  in the $({\bf M},\bar{\bf M})$ representation of $SU(M)^2$, and $N$
dimension $\frac{3}{2}M$ singlets $\tilde{\cal O}_{(j)}$, satisfying an analogous chiral ring relation.
The duality relates the singlet operators of one gauge theory to some of the components of the bi-fundamental operator of the
other gauge theory. 
In either description, the missing components will involve combinations of the singlet operators and instantons.

\subsection{The \texorpdfstring{$+_{N,M,k}$}{monodromy} theory} \label{sec:perp}

As a further ingredient in the construction of 5d superconformal field theories one can add 7-branes to the interior of the 5-brane webs.
These 7-branes source monodromies in the plane of the 5-brane web, and therefore the $(p,q)$ charges of the 5-branes have to be
properly adjusted.
In fact these are not truly new configurations since we can move the 7-branes out of the web and obtain an ordinary 5-brane web.
In so doing additional 5-branes are created, compensating for the removal of the monodromy.
Nevertheless, from the supergravity point of view it is useful to consider the original configuration with the 7-brane in the interior \cite{Gutperle:2018vdd}.

\begin{figure}
\center
\includegraphics[height=0.33\textwidth]{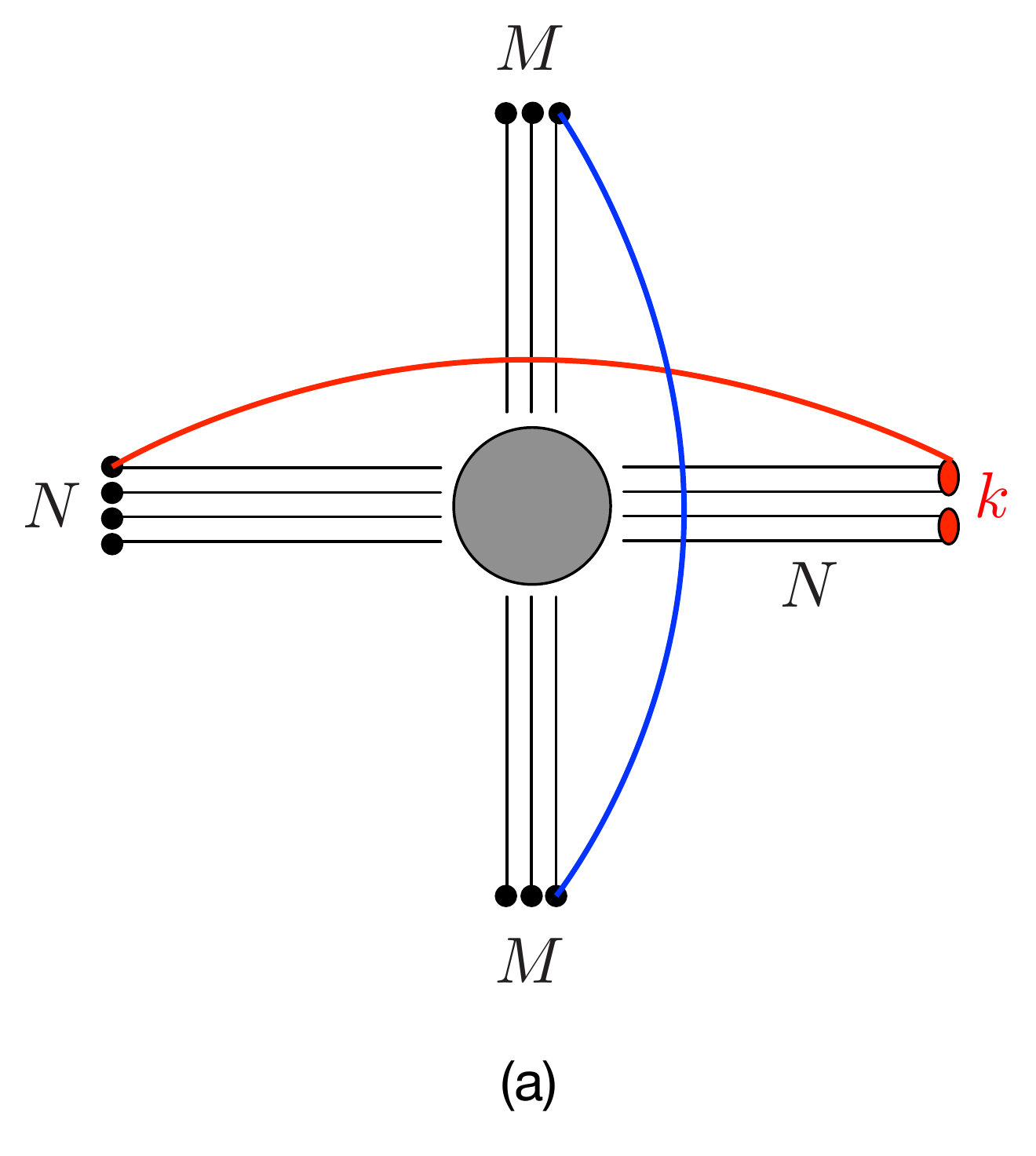} 
\hspace{10pt}
\includegraphics[height=0.33\textwidth]{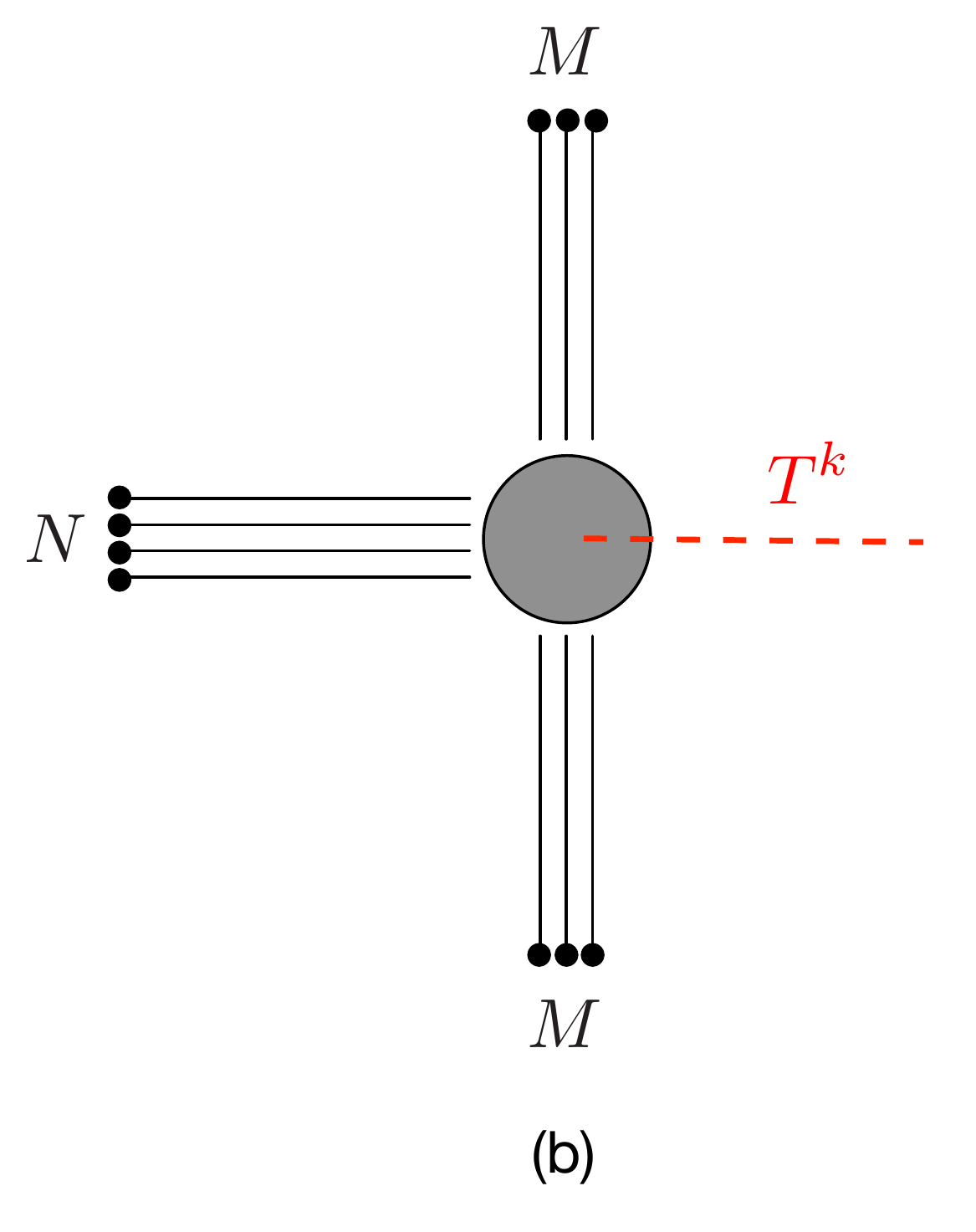} 
\hspace{10pt}
\includegraphics[height=0.28\textwidth]{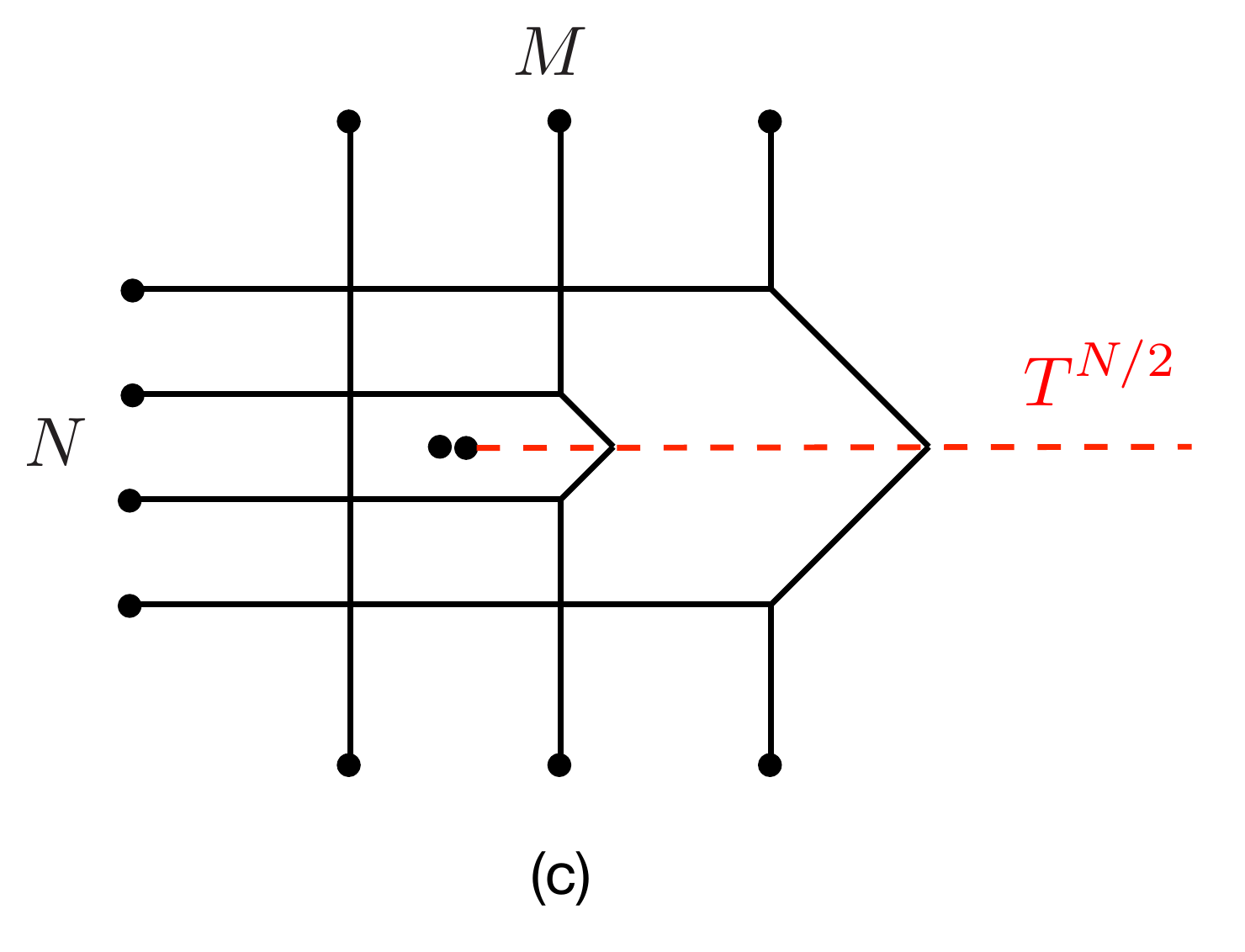} 
\caption{The $+_{N,M,k}$ theory: (a) The $N$ D5-branes on the RHS end on $k$ D7-branes in groups of $N/k$.
(b) An equivalent description with the $k$ D7-branes in the interior. (c) Deformation to an IR quiver gauge theory for $k=N/2$.}
\label{GridNMk}
\end{figure}

As a specific example let us consider the theory described by the 5-brane web of the type shown in fig.~\ref{GridNMk}a.
This theory is a generalization of the $+_{N,M}$ theory discussed in the previous section.
The $N$ D5-branes on the RHS now end on $k$ D7-branes in groups of $\frac{N}{k}$, where $k$ is a divisor of $N$.
We call this the $+_{N,M,k}$ theory.
For $k=N$ this is just the $+_{N,M}$ theory.
More generally it corresponds to the low energy theory along a specific direction on the Higgs branch of the $+_{N,M}$ theory,
in which the global symmetry is reduced to $SU(N)\times SU(k)\times SU(M)^2 \times U(1)$.
The spectrum of chiral operators in this case should include an $({\bf N},\bar{\bf k})$ of $SU(N)\times SU(k)$
coming from strings between D7-branes, and an $({\bf M},\bar{\bf M})$ under $SU(M)^2$ coming from D1-branes
between $(0,1)$ 7-branes.
As we will see below their dimensions are given by 
$\Delta_{({\bf N},\bar{\bf k})} = \frac{3}{2}(M - \frac{N}{k} +1)$ and
$\Delta_{({\bf M},\bar{\bf M})} = \frac{3}{2}N$.
This is consistent with the $k=N$ case, which is just the $+_{N,M}$ theory.
It is also consistent with another special case.
For $k=2$ and $M= \frac{N}{2}+1$ (assuming $N$ is even) this is equivalent to the $Y_{ \frac{N}{2}+1}$ theory
discussed in section \ref{sec:Y-N}.
In this case the $({\bf N},\bar{\bf k}) = ({\bf N},{\bf 2})$ operator has dimension 3, enhancing the global symmetry to
$SU(N+2) \times SU(\frac{N}{2}+1)^2$, consistent with the $Y_{ \frac{N}{2}+1}$ theory.

Now let's move the $k$ D7-branes on the RHS to the interior.
This results in the configuration shown in fig.~\ref{GridNMk}b, which consists of $M$ NS5-branes above and below, 
$N$ D5-branes on the LHS, and a monodromy cut that implements the monodromy $T^k$.
The IR quiver gauge theory given by the deformation corresponding to separating the NS5-branes is given by 
\be
[N]\stackrel{x_1}{-}(N)^{M - \frac{N}{k}-1}\stackrel{x_{M-\frac{N}{k}}}{-}&(N)&\stackrel{x_{M-\frac{N}{k}+1}}{-}(N-k)-(N-2k)-\cdots 
- (2k) \stackrel{x_{M-1}}{-}(k) \nonumber \\
&\;\; | \, {\scriptstyle y}&\\
&[k]& \nonumber
\ee
This has in general a global symmetry $SU(N)\times SU(k)\times U(1)_B^{M} \times U(1)_I^{M - 1}$.
The position of the $k$ flavors in the quiver is determined by the position of the $k$ D7-branes
in the web such that they have no D5-branes attached.
Since one D5-brane is lost for each NS5-brane crossed, the $k$ D7-branes will end up in the $\frac{N}{k}$'th cell from the RHS.
An example with $k=\frac{N}{2}$ is shown in fig.~\ref{GridNMk}c.
For $k=N$ this reduces, as expected, to the IR quiver of the $+_{N,M}$ theory (\ref{GridQuiver1}), 
and for $k=2$ and $M=\frac{N}{2}+1$ it reduces, as expected, to the IR quiver of the $Y_{\frac{N}{2}+1}$ theory (\ref{YNQuiver2}).
In particular in this case we observe that the $SU(N)\times SU(2)$ part of the global symmetry is enhanced together with a $U(1)_B$ to $SU(N+2)$.

The relevant operators in the IR quiver theory include the dimension $\frac{3}{2}(M - \frac{N}{k} +1)$ 
operator 
\be 
{\cal O}^a_{\tilde{b}} = [x_1 \cdots x_{M-\frac{N}{k}} \, y]^a_{\tilde{b}} \,,
\ee
which transforms in the $({\bf N},\bar{\bf k})$ representation of $SU(N)\times SU(k)$ and carries 
$M-\frac{N}{k}+1$ units of charge under the overall $U(1)_B$ symmetry,
and
the dimension $\frac{3}{2} N$ singlets,
\be
{\cal O}_{(i)} &= & \det x_i \\
\tilde{\cal O}_{(j)} &=& \epsilon_{\tilde{b}_1\cdots \tilde{b}_k} \, [(y\, x_{M-\frac{N}{k}+1} \cdots x_{M-j-1})^k]^{\tilde{b}_1\cdots \tilde{b}_k}_{\alpha_1 \cdots \alpha_k} \,
[\det \tilde{x}_{M-j}]^{\alpha_1\cdots \alpha_k} \,,
\ee
where $i = 1, \dots , M-\frac{N}{k}$ and $j=1, \dots , \frac{N}{k}-1$.
These carry $N$ units of overall $U(1)_B$ charge.

\subsection{The \texorpdfstring{$X_{N,M}$}{X-N-M} theory}\label{sec:X-N-M}

\begin{figure}
\center
\includegraphics[height=0.3\textwidth]{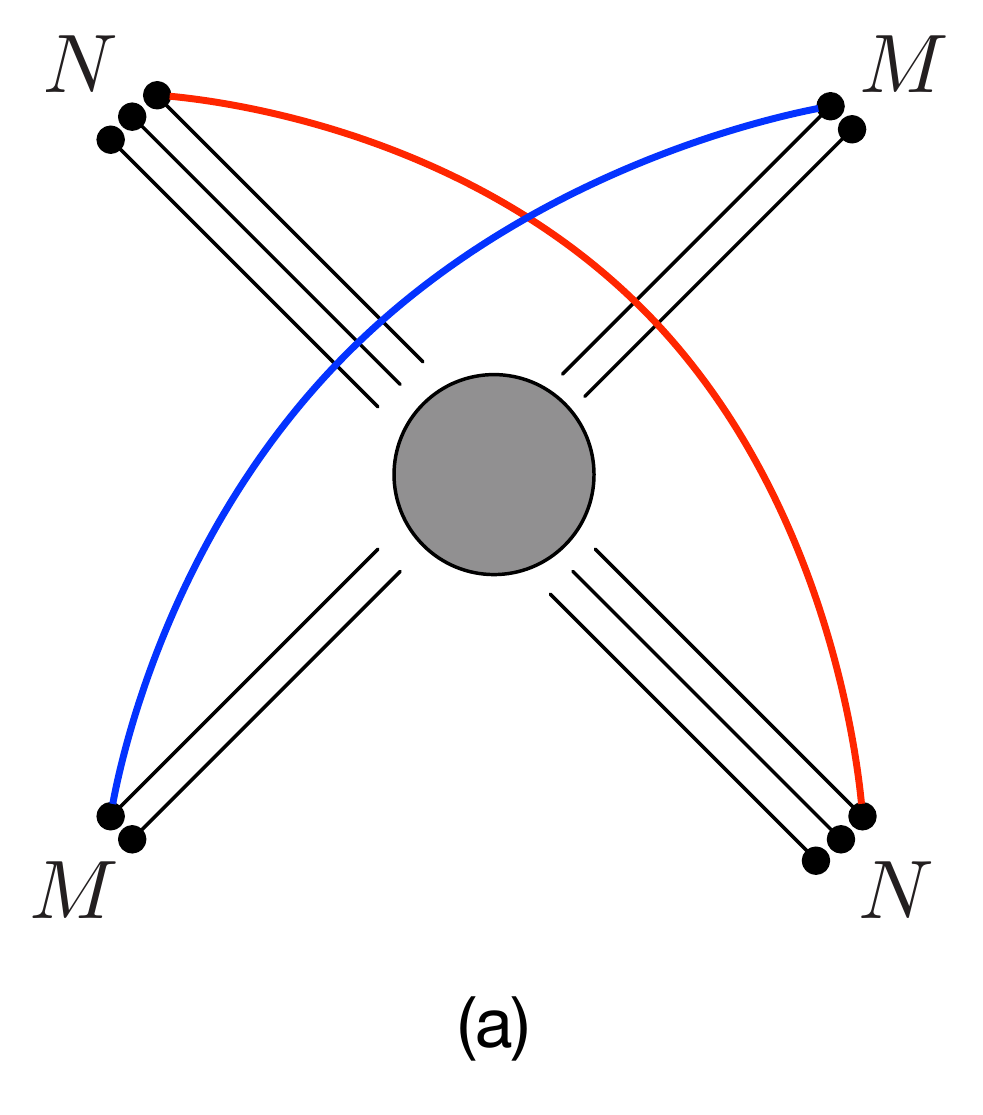} 
\hspace{30pt}
\includegraphics[height=0.3\textwidth]{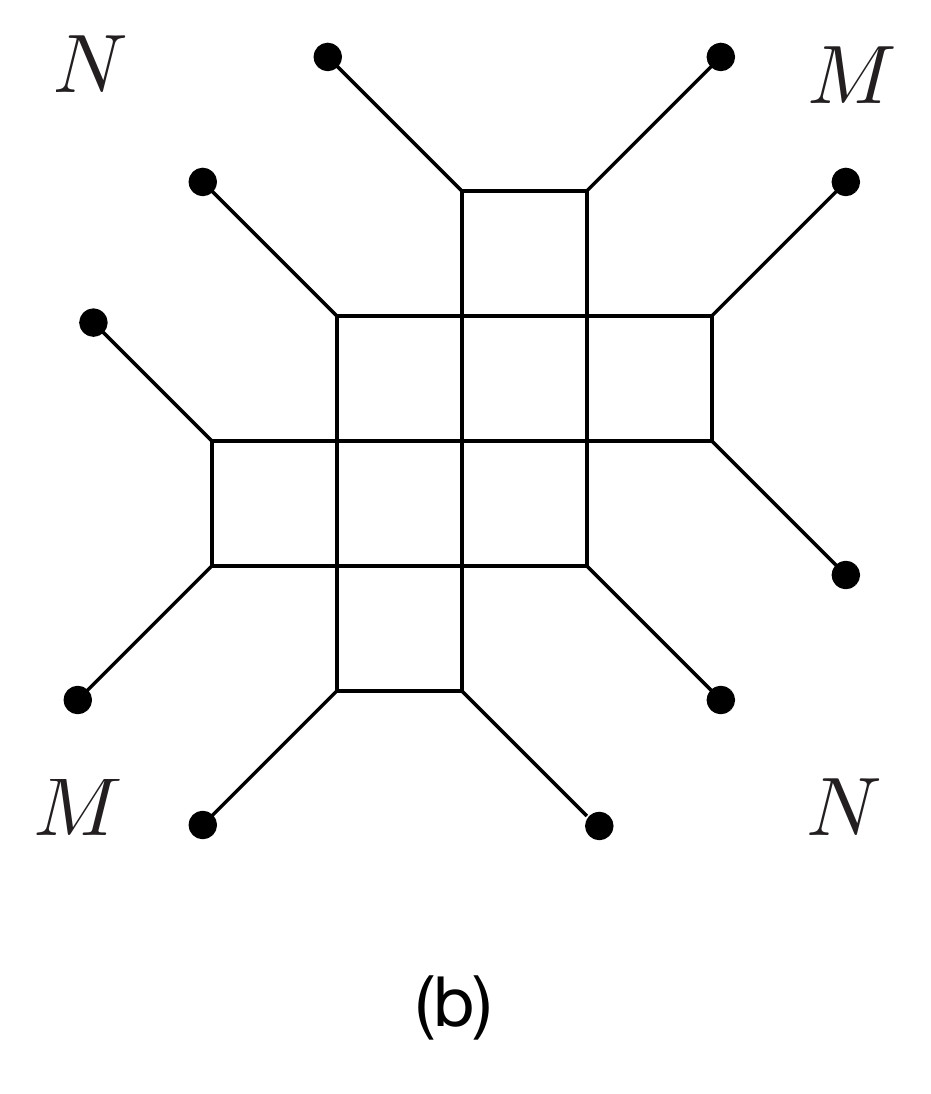} 
\caption{Brane web for the $X_{N_1,N_2}$ theory with $N=3$ and $M=2$ on the left hand side, and a deformation to a quiver gauge theory on the right hand side.}
\label{Xjunction}
\end{figure}

Let us now consider another quartic junction, this time of two sets of $N$ $(1,-1)$5-branes and $M$ $(1,1)$5-branes,
fig~\ref{Xjunction}.
As in the $+_{N,M}$ theory, the global symmetry is generically
$SU(N)^2 \times SU(M)^2 \times U(1)$,
and we expect chiral operators in bi-fundamental representations 
$({\bf N},\bar{\bf N},{\bf 1},{\bf 1})_M$ and
$({\bf 1},{\bf 1},{\bf M},\bar{\bf M})_N$ and their conjugates, corresponding respectively to $(1,1)$ and $(1,-1)$ strings,
as shown in fig.~\ref{Xjunction}(a). 
There is somewhat less direct information in this case about the scaling dimensions of these operators from the IR quiver theories.
However, a number of special cases motivate us to conjecture that these scaling dimensions are given by
$\Delta_{({\bf N},\bar{\bf N})} = 3M$ and $\Delta_{({\bf M},\bar{\bf M})} = 3N$.
For $N=M=1$ this is the $E_1$ theory, which has an $E_1 = SU(2)$ global symmetry.
Our conjecture is consistent with this:
The $U(1)$ symmetry of the $X_{1,1}$ theory is enhanced to $SU(2)$ by these operators.
Note that, as in the $+_{N,M}$ theory, we expect the two bi-fundamental operators to satisfy a chiral ring relation,
that reduces for $N=M=1$ to an equivalence of the two operators.
Furthermore, the theory with $M=1$ and $N=2$ was actually studied in \cite{Bergman:2013aca}, where it was shown that it has a global symmetry $SU(4)$.
This is also consistent with our conjecture.
The $X_{2,1}$ theory has a dimension 3 operator in the $({\bf 2},{\bf 2})_{\pm}$ 
representation of $SU(2)^2 \times U(1)$, enhancing the symmetry to $SU(4)$.

Assuming that $N\geq M$, the IR quiver gauge theory is given by (see fig.~\ref{Xjunction}b)
\be 
(2)\stackrel{x_1}{-}(4)-\cdots-(2M-2)\stackrel{x_{M-1}}{-}(2M)^{N-M+1}\stackrel{x_{N}}{-}(2M-2)-\cdots - (4) \stackrel{x_{N+M-2}}{-}(2) \,.\nonumber \\
\ee
The global symmetry is $U(1)_B^{N+M-2} \times U(1)_I^{N+M-1}$.
The S-dual quiver theory is the same, and therefore the quiver theories for $X_{N,M}$ and $X_{M,N}$ are the same.
For the two special cases we considered above, $(N,M)=(1,1)$ and $(2,1)$, the IR gauge theory is the pure $SU(2)_0$ theory
and the $SU(2)_0\times SU(2)_0$ quiver, where the subscripts denote the values of the discrete theta parameters.
The former is the IR deformation of the $E_1$ theory, and the latter is one of the theories studied in \cite{Bergman:2013aca}.
The IR gauge theory has dimension $3M$ operators given by 
\be
{\cal O}_i= \det x_i \,,
\ee
where $i=M,\ldots , N-1$.
These are presumably some of the components of the $({\bf N},\bar{\bf N})$ operator.

\begin{figure}
\center
\includegraphics{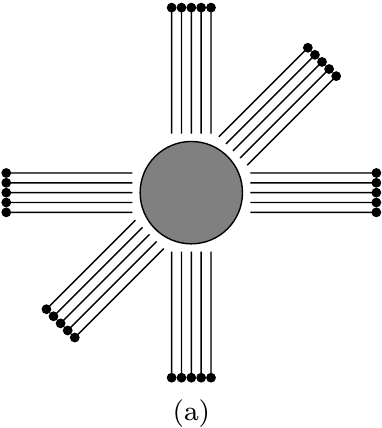} 
\hspace{30pt}
 \includegraphics{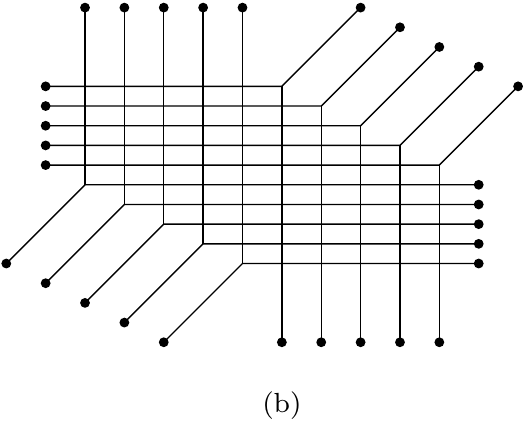}
\caption{Brane web for the $\pslash_N$ theory with $N=5$ on the left hand side, and a deformation to a quiver theory on the right.}
\label{fig:pslash}
\end{figure}


\subsection{The \texorpdfstring{$\pslash_N$}{pslash} theory}\label{sec:pslash}

As a last example we consider the sextic junction with $N$ D5 branes, $N$ NS5 branes and $N$ (1,1)5-branes, as shown in fig.~\ref{fig:pslash}. 
The global symmetry is in general $SU(N)^6\times U(1)^3$. 
We expect the spectrum of chiral operators to include the bi-fundamentals
$({\bf N},\bar{\bf N},{\bf 1},{\bf 1},{\bf 1},{\bf 1})$,
$({\bf 1},{\bf 1},{\bf N},\bar{\bf N},{\bf 1},{\bf 1})$, and
$({\bf 1},{\bf 1},{\bf 1},{\bf 1},{\bf N},\bar{\bf N})$,
corresponding, respectively, to $(1,0)$, $(0,1)$, and $(1,1)$ strings.
We also expect operators transforming in the 
$({\bf N},{\bf 1},{\bf N},{\bf 1},{\bf N},{\bf 1})$ and $({\bf 1},{\bf N},{\bf 1},{\bf N},{\bf 1},{\bf N})$
corresponding to 3-pronged strings, as in the $T_N$ theory.
Below we will identify a subset of these operators in the IR quiver gauge theory.
In particular we will see that their scaling dimensions are given by 
$\Delta_{({\bf N},\bar{\bf N})} = 3N$, and $\Delta_{({\bf N},{\bf N},{\bf N})} = \frac{3}{2}(3N-1)$.
This is also consistent with the simplest case of $N=1$, which is the $E_3$ theory.
In this theory the global symmetry is enhanced from $U(1)^3$ to $E_3 = SU(3)\times SU(2)$,
where some of the additional conserved currents are provided by the bi-fundamental and tri-fundamental operators, all of which have dimension 3.

The IR quiver gauge theory (as well as the S-dual quiver theory) is given by (fig.~\ref{fig:pslash})
\begin{align}
 [N]\stackrel{x_1}{-}(N+1)\stackrel{x_2}{-}\dots - (2N-1)\stackrel{x_N}{-}(2N)\stackrel{x_{N+1}}{-}(2N-1)-\dots -(N+1)\stackrel{x_{2N}}{-}[N]~.
\end{align}
This exhibits a global symmetry $SU(N)^2\times U(1)_B^{2N} \times U(1)_I^{2N-1}$.
There is a dimension $3N$ operator
in the $({\bf N},\bar{\bf N})$ representation given by
\be
{\cal O}^a_{\tilde{b}} = [x_1 \cdots x_{2N}]^a_{\tilde{b}} \,,
\ee
and $N$ dimension $\frac{3}{2}(3N-1)$ operators in the $({\bf N},{\bf 1})$ and $({\bf 1},{\bf N})$ representations
\be
{\cal O}_{\tilde{b}(i)} &=& [\det \tilde{x}_i]_{\alpha} \, [x_{i+1} \cdots x_{2N}]^{\alpha}_{\tilde{b}}  \\
{\cal O}_{(i)}^{a} &=&  [x_1\cdots x_{N+i-1}]^{a}_{\beta} \, [\det \tilde{x}_{N+i}]^{\beta} 
\ee
where $i=1, \ldots , N$.


\section{Type IIB warped \texorpdfstring{$AdS_6$}{AdS6} solutions}\label{sec:sugra-review}

This section contains 
an overview of the warped $AdS_6\times S^2\times\Sigma$ supergravity solutions constructed in \cite{DHoker:2016ujz,DHoker:2016ysh,DHoker:2017mds,DHoker:2017zwj}, to introduce the relevant notation. We discuss the near-pole behavior and the normalization of the 5-brane charges encoded in the supergravity solutions. We identify, in parts, a fluctuation which is universally present for all solutions and whose $AdS_6$ part is a massless gauge field. We also identify the gauge fields dual to the $U(1)$ factors in (\ref{eq:global-symmetry}).

\subsection{Review of the solutions}

The geometry in the solutions of \cite{DHoker:2016ujz,DHoker:2016ysh,DHoker:2017mds,DHoker:2017zwj} is a warped product of $AdS_6$ and $S^2$ over a Riemann surface $\Sigma$.
The metric and two-form field are parametrized in terms of functions $f_6^2$, $f_2^2$, $\rho^2$ and $\cC$ on $\Sigma$ as follows
\begin{align}\label{eqn:ansatz}
 ds^2 &= f_6^2 \, ds^2 _{\mathrm{AdS}_6} + f_2 ^2 \, ds^2 _{\mathrm{S}^2} + 4\rho^2 |dw|^2~,
 &
 C_{(2)}&=\cC \vol_{S^2}~.
\end{align}
The solutions to the BPS equations for preserving 16 supersymmetries are expressed in terms of two locally holomorphic functions $\cA_\pm$ on $\Sigma$, from which we define the composite quantities
\begin{align}
 \kappa^2&=-|\partial_w \cA_+|^2+|\partial_w \cA_-|^2~,
 &
 \partial_w\cB&=\cA_+\partial_w \cA_- - \cA_-\partial_w\cA_+~,
 \nonumber\\
 \cG&=|\cA_+|^2-|\cA_-|^2+\cB+\bar{\cB}~,
 &
  R+\frac{1}{R}&=2+6\,\frac{\kappa^2 \, \cG }{|\partial_w\cG|^2}~.
 \label{eq:kappa-G}
\end{align}
The metric functions are then given by
\begin{align}\label{eqn:metric}
f_6^2&=\sqrt{6\cG} \left ( \frac{1+R}{1-R} \right ) ^{\tfrac{1}{2}},
&
f_2^2&=\frac{1}{9}\sqrt{6\cG} \left ( \frac{1-R}{1+R} \right ) ^{\tfrac{3}{2}},
&
\rho^2&=\frac{\kappa^2}{\sqrt{6\cG}} \left (\frac{1+R}{1-R} \right ) ^{\tfrac{1}{2}},
\end{align}
while the axion-dilaton scalar is parametrized as follows
\begin{align}\label{eq:B-def}
B &=\frac{\partial_w \cA_+ \,  \partial_{\bar w} \cG - R \, \partial_{\bar w} \bar \cA_-   \partial_w \cG}{
R \, \partial_{\bar w}  \bar \cA_+ \partial_w \cG - \partial_w \cA_- \partial_{\bar w}  \cG}~,
&
B&=\frac{1+i\tau}{1-i\tau}~,
&
\tau&=\chi+i e^{-2\phi}~.
\end{align}
Note the normalization convention for the dilaton.
Finally, the complex function $\cC$ parametrizing the two-form potential is given by
\begin{align}\label{eqn:flux}
 \cC = \frac{4 i }{9}\left (  
\frac{\partial_{\bar w} \bar \cA_- \, \partial_w \cG}{\kappa ^2} 
- 2 R \, \frac{  \partial_w \cG \, \partial_{\bar w} \bar \cA_- +  \partial_{\bar w}  \cG \, \partial_w \cA_+}{(R+1)^2 \, \kappa^2 }  
 - \bar  \cA_- - 2 \cA_+ \right )~.
\end{align}
To obtain physically regular solutions, additional regularity conditions have to be implemented. This was carried out for solutions without monodromy in \cite{DHoker:2017mds}, and extended to solutions with monodromy in \cite{DHoker:2017zwj}. The Riemann surface $\Sigma$ is taken as the disc, or equivalently the upper half plane, in both cases.

\subsubsection{Solutions without monodromy}

With a complex coordinate $w$ on the upper half plane, the functions $\cA_\pm$ for regular solutions without monodromy are given by
\begin{align}\label{eqn:cA}
 \cA_\pm (w) &=\cA_\pm^0+\sum_{\ell=1}^L Z_\pm^\ell \ln(w-r_\ell)~,
 &
 \overline{\cA_\pm^0}&=-\cA_\mp^0~.
\end{align}
The $r_\ell$ are the locations of poles in the differentials $\partial_w\cA_\pm$ on the real line, and the residues are given by
\begin{align}\label{eqn:residues}
Z_+^\ell  &=
 \sigma\prod_{n=1}^{L-2}(r_\ell-s_n)\prod_{k \neq\ell}^L\frac{1}{r_\ell-r_k}~,
 &
 Z_-^\ell&=-\overline{Z_+^\ell}~,
\end{align}
with an overall complex normalization $\sigma$ and complex parameters $s_n$ inside $\Sigma$.
Regularity imposes further constraints on these parameters. With $Z^{[\ell k]}\equiv Z_+^\ell Z_-^k-Z_+^k Z_-^\ell$, they are given by
\begin{align}
\label{eqn:constr}
 \cA_+^0 Z_-^k - \cA_-^0 Z_+^k 
+ \sum _{\ell \not= k }Z^{[\ell k]} \ln |r_\ell - r_k| &=0~,
&k&=1,\cdots,L~.
\end{align}

\subsubsection{Solutions with D7-brane monodromy}
The additional parameters for a solution with D7-brane punctures are the locations of the punctures in $\Sigma$, $w_i$, a real number $n_i$ for each puncture and a phase $\gamma_i$ specifying the orientation of the associated branch cut. The functions $\cA_\pm$ are given by
\begin{align}\label{eqn:cA-monodromy}
 \cA_\pm&= \cA_\pm^0+\sum_{\ell=1}^L Z_\pm^\ell \ln(w-r_\ell) + \int_\infty^w dz \;f(z)\sum_{\ell=1}^L \frac{Y^\ell}{z-r_\ell}~,
\end{align}
with $Y^\ell\equiv Z_+^\ell-Z_-^\ell$, the constants related by $\cA^0_+=-\bar\cA_-^0$, and
\begin{align}
f(w) &= \sum _{i=1}^I \frac{n_i^2}{4\pi} \ln \left ( \gamma_i\,\frac{ w-w_i}{w -\bar w_i} \right )~.
\end{align}
The contour for the integration in (\ref{eqn:cA-monodromy}) is chosen such that no branch cuts are crossed.
The regularity constraints that the parameters have to satisfy are
\begin{align}
\label{eq:w1-summary}
 0&=2\cA_+^0-2\cA_-^0+\sum_{\ell=1}^LY^\ell \ln|w_i-r_\ell|^2~,
 &i&=1,\cdots,I~,
 \\
 0&=2\cA_+^0\cY_-^k-2\cA_-^0\cY_+^k
 +\sum_{\ell\neq k} Z^{[\ell, k]}\ln |r_\ell-r_k|^2+Y^kJ_k~,
 &
 k&=1,\cdots,L~.
 \label{eq:DeltaG0-summary}
\end{align}
With $\cS_k$ denoting the set of branch points for which the associated branch cut intersects the real line in $(r_k,\infty)$, $J_k$ is given by
\begin{align}\label{eq:Jk}
 J_k&=\sum_{\ell=1}^L Y^\ell\Bigg[\int_\infty^{r_k} dx f^\prime(x)  \ln |x-r_\ell|^2
 +\sum_{i\in\cS_k} \frac{i n_i^2}{2} \ln |w_i-r_\ell|^2\Bigg]~.
\end{align}
The residues of the differentials of (\ref{eqn:cA-monodromy}) at the poles $r_\ell$ are given by
\begin{align}\label{eq:cY}
\cY_\pm^\ell&=Z_\pm^\ell+ f(r_\ell)Y^\ell~.
\end{align}

\subsection{Behavior near poles and punctures}
The behavior of the supergravity fields near poles and punctures has been derived in \cite{DHoker:2017mds,DHoker:2017zwj}. We will collect the expressions here for convenience and extend the discussion to poles with purely imaginary residues. We denote the residues at the poles by $Z_\pm^\ell$, which is the appropriate notation for solutions without monodromy. For solutions with monodromy, the expressions take the same form, but with $Z_\pm^\ell$ replaced by $\cY_\pm^\ell$.

The $SL(2,\RR)$ invariant Einstein-frame metric functions near a pole $r_m$, expressed in radial coordinates centered on the pole, $w=r_m+r e^{i\theta}$, are given by
\begin{align}\label{eq:Einst-metric-exp}
 f_6^2&\approx 2\cdot 3^{\tfrac{1}{4}}\kappa_m r^{\tfrac{1}{2}}|\ln r|^{\tfrac{3}{4}}~,
 &
 \rho^2&\approx \frac{\kappa_m}{2\cdot 3^{3/4}} r^{-\tfrac{3}{2}}|\ln r|^{-\tfrac{1}{4}}~,
 &
 f_2^2&\approx 4 r^2\sin^2\!\theta\,\rho^2~,
\end{align}
with a constant $\kappa_m$.
With $|dw|^2=dr^2+d\theta^2$ in (\ref{eqn:ansatz}), the $S^2$ combines with the $d\theta^2$ term to form an $S^3$ around the pole.
The complex three-form field strength $F_{(3)}=dC_{(2)}$ near the pole is given by
\begin{align}\label{eq:3-form-near-pole}
 F_{(3)}&=\frac{8}{3}Z_+^m \vol_{S^3}~, 
 & \vol_{S^3}&=\sin^2\!\theta\,d\theta\wedge\vol_{S^2}~.
\end{align}
For poles with a non-vanishing real part of the residue, the dilaton and axion are given by
\begin{align}\label{eq:axion-dilaton-near-pole}
 e^{- 2 \phi} & \approx  \frac{ \sqrt{3}  \, \kappa_m^2}{|Z_+^m-Z_-^m|^2} \,  r \, |\ln r | ^{-\frac{1}{2}}~,
 &
 \chi & \approx  i \, \frac{ Z_-^m +  Z_+^m}{Z_-^m -  Z_+^m}~.
\end{align}
For poles with a purely imaginary residue, the corresponding expressions can be obtained from an S-duality transformation.
For a purely real residue, such that the pole corresponds to NS5 branes, $\chi\approx 0$.
An $SU(1,1)$ transformation, as defined in sec.~2.2 of \cite{DHoker:2017mds}, with $u=i$ and $v=0$, transforms a pole with a real residue to a pole with an imaginary residue.
For a vanishing axion this transformation acts as $\phi\rightarrow \phi^\prime=-\phi$, while $\chi^\prime=\chi=0$. We thus find
\begin{align}\label{eq:chi-phi-D5}
 e^{+2\phi}&\approx\frac{\sqrt{3}\kappa_m^2}{4|Z_+^m|^2}r|\ln r|^{-1/2}~,
 &
 \chi&\approx 0~,
 & \text{for \ $Z_+^m=Z_-^m$.}
\end{align}

For solutions with monodromy, the expressions for the near-pole solution hold with $Z_\pm^m$ replaced by $\cY_\pm^m$. Near a puncture $w_i$ with a D7-brane monodromy, the  metric and axion-dilaton scalar $\tau$ are given in terms of a local coordinate $z$ with the puncture at $z=0$ by
\begin{align}\label{eq:near-puncture}
ds^2 &\approx
ds^2_{AdS_6\times S^2}
+\Im(\cH) |dz |^2 ~,
&
\tau &\approx\cH +\tilde\tau_0~,
&
\cH=-\frac{in_i^2}{2\pi}\ln z~.
\end{align}

\subsection{5-brane and 7-brane charges}\label{sec:asympt-charge}

The poles were identified with $(p,q)$ 5-branes in \cite{DHoker:2017mds,DHoker:2017zwj}, with the real/imaginary part of the residue related to the NS5/D5 charge, and the punctures were identified with 7-branes. We now discuss the normalization of the 
5-brane and 7-brane charges. 

The complex two-form $C_{(2)}$ splits into real and imaginary parts, corresponding to the NS-NS two-form field $B_2$ and the R-R two-form potential $C_{(2)}^{\rm RR}$, 
\begin{align}
 C_{(2)}&=B_2+i C_{(2)}^{\rm RR}~.
\end{align}
The charge quantization conditions are derived from the coupling of fundamental strings and D1-branes to $B_2$ and $C_{(2)}^{\rm RR}$, respectively. For the normalization of the supergravity action and brane tensions we follow the conventions of \cite{deAlwis:1997gq}. A fundamental string couples to  $B_2$ through $-T\int B_2$,
and the Dirac quantization condition yields
\begin{align}
 T\int_{S^3} dB_2 &=2\pi N_{\rm NS5}~,
 &
 T&=\frac{1}{2\pi\alpha^\prime}~,
\end{align}
with an integer $N_{\rm NS5}$.  Taking the $S^3$ formed around the pole $r_m$, and using the near-pole behavior of $F_{(3)}=dC_{(2)}$ in (\ref{eq:3-form-near-pole}), yields
\begin{align}
 \Re(Z_+^m)&=\frac{3}{4}\alpha^\prime N_{\rm NS5}~,
\end{align}
where $N_{\rm NS5}$ is the number of NS5 branes at the pole. 
Simliarly, the coupling of D1-branes to $C_{(2)}^{\rm RR}$ yields $\Im(Z_+^m)=\frac{3}{4}\alpha^\prime N_{\rm D5}$, and thus
\begin{align}\label{eq:residue-N}
 Z_+^m&=\frac{3}{4}\alpha^\prime (N_{\rm NS5}+i N_{\rm D5})~.
\end{align}
When referring to $(p,q)$ 5-branes we use $p=N_{\rm D5}$ and $q=N_{\rm NS5}$, such that a $(1,0)$ 5-brane corresponds to one D5-brane and a $(0,1)$ 5-brane to one NS5-brane.

For solutions with monodromy, (\ref{eq:residue-N}) holds with $Z_+^m$ replaced by $\cY_+^m$.
For D7-brane punctures, the monodromy of the axion-dilaton scalar as given in (\ref{eq:near-puncture}) around $z=0$  is  $\tau\rightarrow \tau+n_i^2$. Since $\tau\rightarrow \tau+1$ for a single D7 brane, we conlude that the number of D7-branes at the puncture $w_i$ is given by
\begin{align}
 N_{\rm D7}&=n_i^2~.
\end{align}

\subsection{\texorpdfstring{$R$}{R}-symmetry gauge field fluctuation}\label{sec:gauge-field}

The $S^2$ factor in the geometry of the $AdS_6\times S^2\times\Sigma$ solutions, with its corresponding isometries, geometrically realizes the $R$-symmetry of the dual SCFTs. It suggests that there should be a fluctuation around generic $AdS_6\times S^2\times\Sigma$ solutions which corresponds to a massless $SU(2)$ gauge field on $AdS_6$, holographically dual to the conserved $R$-symmetry current in the SCFT. 
This fluctuation is part of the multiplet including the metric fluctuation dual to the energy-momentum tensor, identified recently in \cite{Gutperle:2018wuk}. In this section we identify the parts of the gauge field fluctuation that will be relevant for determining the $R$-symmetry charges of the string states to be discussed in the next section.

With $K_I$, $I=1,2,3$ denoting the Killing vector fields on a unit radius $S^2$ and $A^I$ a set of one-forms on $AdS_6$, the general ansatz for the metric perturbation in the spirit of non-Abelian Kaluza-Klein reduction\footnote{A review can be found in \cite{Duff:1986hr} and an interesting historical note in \cite{Straumann:2000zc}.} is obtained by replacing
\begin{align}\label{eq:SU2-pert}
 dx^\mu&\rightarrow dx^\mu+ K_I^\mu A^I~.
\end{align}
The combination on the right hand side is invariant under the linearized gauge transformations
\begin{align}\label{eq:gauge-trafo}
 \delta x^\mu&= -K_I^\mu\lambda^I~, & \delta A^I&=d\lambda^I~,
\end{align}
where $\lambda^I$ is a set of functions on $AdS_6$ that are of the same order as $A^I$ in the (implicit) small parameter characterizing the perturbative expansion.
With an explicit parametrization of the $S^2$ in the metric (\ref{eqn:ansatz}) as
\begin{align}\label{eq:S2-expl}
 ds^2_{S^2}&=d\theta_1^2+\sin^2\!\theta_1\,d\theta_2^2~,
\end{align}
a basis for the $S^2$ Killing vector fields in the full $AdS_6\times S^2\times \Sigma$ spacetime is given by
\begin{align}
\label{K's}
 K_1&=\sin\theta_2\partial_{\theta_1}+\cot\theta_1\cos\theta_2\partial_{\theta_2}~,
 \nonumber\\
 K_2&=\cos\theta_2\partial_{\theta_1}-\cot\theta_1\sin\theta_2\partial_{\theta_2}~,
 &
 K_3&=\partial_{\theta_2}~.
\end{align}
The dual one-forms $\tilde K_I=(g_{S^2})_{\mu \nu}K_I^\mu dx^\nu$ satisfy $d\star_{S^2} \tilde K_I=0$, so we can introduce functions $f_I$ with $\star_{S^2}\tilde K_I=df_I$. They are given by
\begin{align}\label{eq:fI}
 f_1&=-\sin\theta_1\cos\theta_2~,
 &
 f_2&=\sin\theta_1\sin\theta_2~,
 &
 f_3&=\cos\theta_1~.
\end{align}
The perturbation to the metric (\ref{eqn:ansatz}) resulting from (\ref{eq:SU2-pert}) then takes the form
\begin{align}
 ds^2_{S^2}&\rightarrow \left(d\theta_1+K_I^{\theta_1}A^I\right)^2+\sin^2\!\theta_1\,\left(d\theta_2+K_I^{\theta_2}A^I\right)^2~.
\end{align}
Provided that $A^I$ satisfies the equation of motion for a massless gauge field on $AdS_6$, 
\begin{align}
 0&=\hat\nabla^m (dA)_{mn}~,
\end{align}
where $\hat\nabla^m$ denotes the canonical covariant derivative on unit radius $AdS_6$,
the perturbed Ricci tensor takes a particularly simple form. Namely, its components in the perturbed frame
\begin{align}\label{eq:pert-frame}
 \tilde e^A&=e^A+\delta e^A~, & \delta e^A&=K_I^A A^I~,
\end{align}
are identical to the components of the unperturbed Ricci tensor in the unperturbed frame. This facilitates solving the perturbed Einstein's equations. 

In order to solve the full linearized equations of motion, this ansatz has to be extended to the entire set of bosonic supergravity fields. For the warped product geometries considered here, this entails a proper treatment of the dependence on $\Sigma$ and of the non-trivial background values for the remaining supergravity fields. In particular, the complex three-form field strength given by
\begin{align}\label{eq:bg-F3}
 F_{(3)}&=d\cC\wedge \vol_{S^2}~,
\end{align}
is by itself not invariant under the infinitesimal gauge transformations (\ref{eq:gauge-trafo}).
The replacement in (\ref{eq:SU2-pert}) indeed acts non-trivially on the volume form, resulting in
\begin{align}
 \vol_{S^2} &\rightarrow {\vol_{S^2}}-df_I\wedge A^I~.
 \label{eq:delta-volS2}
\end{align}
This restores invariance of (\ref{eq:bg-F3}) under the gauge transformations (\ref{eq:gauge-trafo}), but the result is not closed and further terms are required. Invariance under the gauge transformations (\ref{eq:gauge-trafo}) together with the Bianchi identity for $F_{(3)}$ implies that the perturbation to the two-form potential takes the form
\begin{align}\label{eq:delta-C2}
 C_{(2)}+\delta C_{(2)}&=\cC{\vol_{S^2}} + d\cC \wedge f_I A^I+\ldots~,
\end{align}
where the dots denote terms involving $A^I$ only through the linearized field strength $F^I=dA^I$.
The complete perturbation may also involve a non-trivial 4-form potential $\delta C_{(4)}$: while the symmetry of the background solution was sufficient to fix $C_{(4)}=0$, this is not the case anymore in the perturbed configuration.
We will leave a complete discussion of the linearized equations of motion and their solution for the future.
The part that will be relevant in the next section is the term explicitly involving $A^I$ in (\ref{eq:delta-C2}), whose form is fixed by invariance under the linearized gauge transformations (\ref{eq:gauge-trafo}).


\subsection{Gauge fields from 3-cycles}\label{sec:3-cycles}

\begin{figure}
\center
\includegraphics[height=0.28\textwidth]{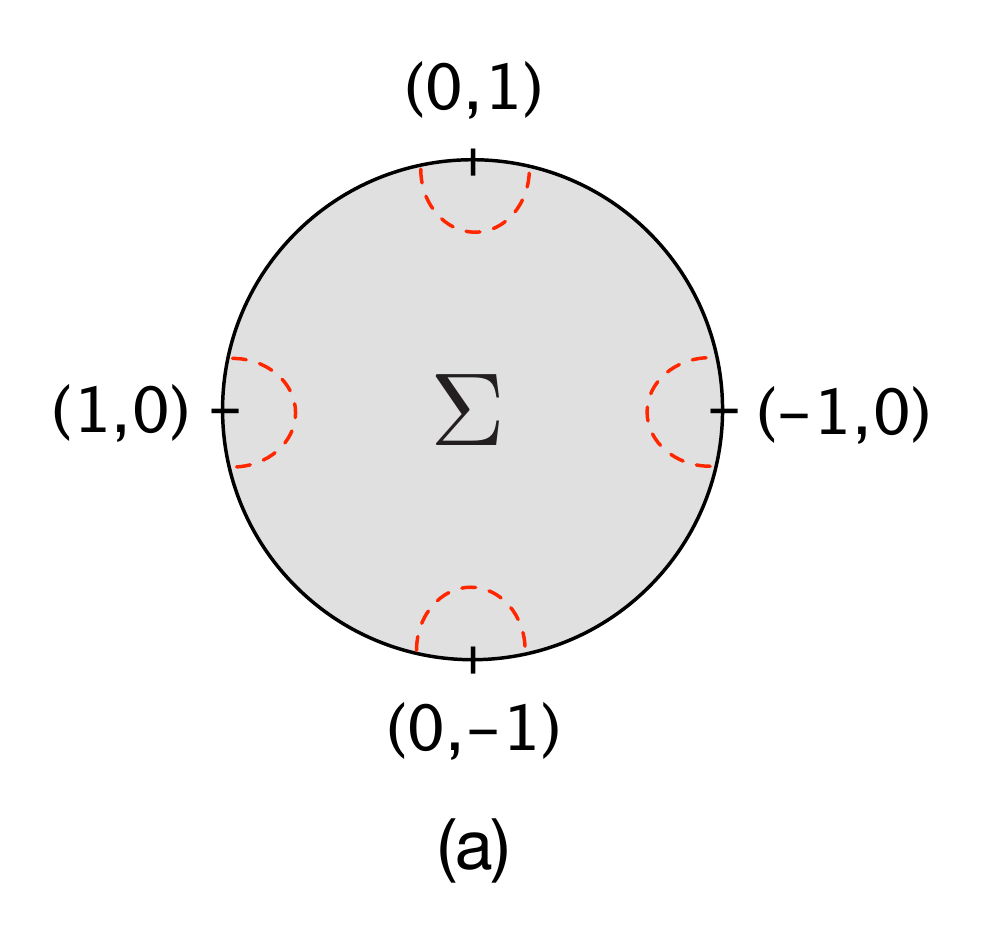} 
\hspace{10pt}
\includegraphics[height=0.28\textwidth]{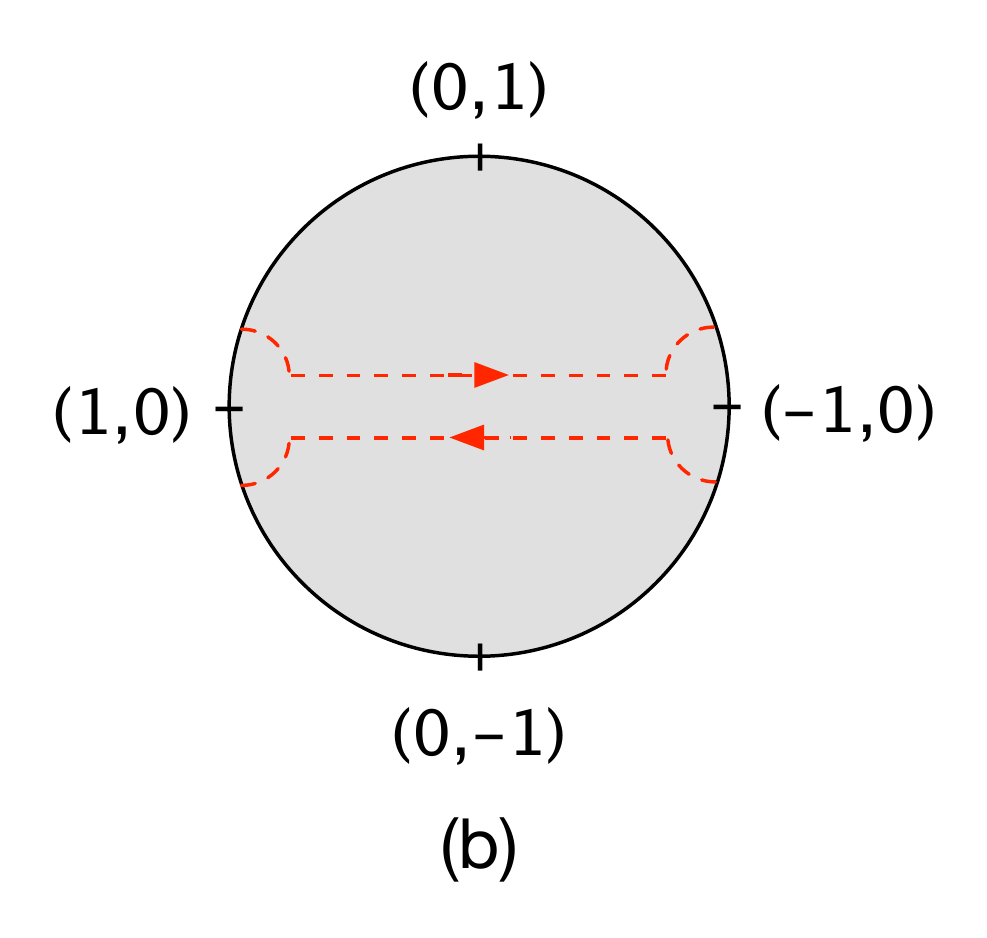} 
\hspace{10pt}
\includegraphics[height=0.28\textwidth]{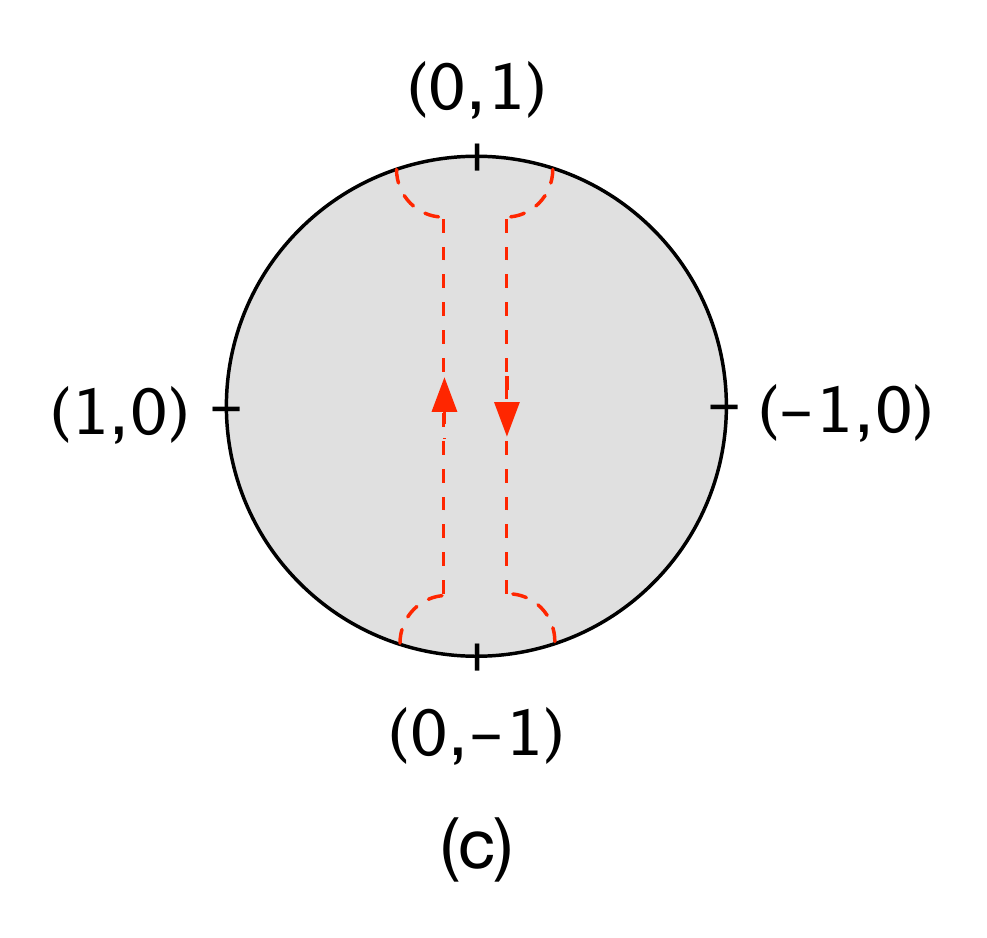} 
\caption{3-cycles in the solution corresponding to $(\pm 1,0)$ and $(0,\pm 1)$ 5-branes.
(a) The basis cycles $c_{(\pm 1,0)}$, $c_{(0,\pm 1)}$ carry nonzero flux. (b) The cycle $c_{(1,0)} + c_{(-1,0)}$ has a vanishing flux.
(c) The cycle $c_{(0,1)} + c_{(0,-1)}$ also has a vanishing flux, but it is equivalent to the cycle $c_{(1,0)} + c_{(-1,0)}$.}
\label{ThreeCycles}
\end{figure}

Recall that to each pole on the boundary of $\Sigma$ there is an associated 3-cycle.
This 3-cycle is given by fibering the $S^2$
over a curve that starts at the boundary of $\Sigma$ on one side of the pole and ends at the boundary
of $\Sigma$ on the other side of the pole.
Since the reduction of the RR 4-form potential over such a cycle gives a 1-form in $AdS_6$,
we seem to obtain $L$ massless vector fields, {\em i.e.} $U(1)$ gauge fields.
However, this overcounts the number of massless vector fields in two ways.
First, the sum of the $L$ cycles is a trivial cycle since, if there are no punctures in the interior of $\Sigma$, 
it can be contracted to a point inside $\Sigma$.
So there are $L-1$ nontrivial 3-cycles.
Second, the fields obtained in the reduction are not all massless.
Cycles with a non-vanishing complex 3-form flux $F_{(3)}$ lead to a massive, or more generally gapped, $U(1)$ gauge field.
This can be seen indirectly by considering a charged particle, which is described in the Type IIB picture by 
a D3-brane wrapping the corresponding 3-cycle.
If the the total $F_{(3)}$ flux is non-vanishing the D3-brane has a tadpole that requires attaching strings of the appropriate
type to it. The charged particle will therefore come with strings going to the boundary of $AdS_6$,
signaling that the corresponding gauge field is confined.
The number of massless $U(1)$ gauge fields is given by the number of linearly independent 3-cycles 
on which $F_{(3)}=0$.
Let us denote the basis of 3-cycles corresponding to the poles by $\{c_\ell\}$, where $\ell = 1,\ldots, L$,
and $\sum_\ell c_\ell = 0$.
The cycles with vanishing flux are then given by $\sum_\ell a_\ell c_\ell$, where $a_\ell$ are non-negative integers satisfying
$\sum_\ell a_\ell Z_{\pm}^\ell= 0$.
The space of solutions is $L-2$ dimensional, but the condition $\sum_\ell c_\ell = 0$ reduces it to $L-3$.
We therefore get $L-3$ massless $U(1)$ gauge fields, in agreement with the number of $U(1)$ factors
in the global symmetry (\ref{eq:global-symmetry}).
We illustrate this in fig.~\ref{ThreeCycles} for an example with $L=4$ corresponding to $(\pm 1,0)$ 5-branes and $(0,\pm 1)$ 5-branes.

\section{BPS states from supergravity}\label{sec:strings-in-warped-AdS6}

In this section we realize supergravity solutions that are expected to describe the large-$N$ limits of the field theories discussed in sec.~\ref{sec:5dscft}, and study $(p,q)$ strings and string-webs embedded into the solutions. Using the gauge field fluctuation discussed in sec.~\ref{sec:gauge-field}, we identify BPS states and compute the scaling dimensions of the corresponding dual operators from the supergravity description.

\subsection{\texorpdfstring{$(p,q)$}{(p,q)} strings in warped \texorpdfstring{$AdS_6$}{AdS6}}\label{sec:D1F1}
In this section we discuss the general features of $(p,q)$-string embeddings into the warped $AdS_6$ solutions.
For the unit-radius $AdS_6$ factor of the background geometry we use global coordinates
\begin{align}\label{eq:AdS6-global}
 ds^2_{AdS_6}&=\frac{du^2}{1+u^2} - (1+u^2)dt^2+u^2ds^2_{S^4}~,
\end{align}
such that the dual SCFT is realized on $\RR\times S^4$.
We seek static configurations where the embedding wraps the time direction, $t$, in $AdS_6$ and a one-dimensional subspace of $\Sigma$, which we parametrize by a real coordinate $\xi$. We fix the spatial position in $AdS_6$ to $u=0$, such that an entire $\rm SO(5)$ subgroup of the spatial isometries in $\rm SO(2,5)$ is preserved along with time translations. 
This corresponds in the SCFT to an operator insertion at the origin in radial quantization, and the on-shell Hamiltonian of the $(p,q)$ strings is then related to the scaling dimension of the dual operator. We denote the metric induced by the Einstein-frame metric on the worldvolume by $g$, and for the class of embeddings discussed here it is given by
\begin{align}
 g&=-f_6^2 dt^2+4\rho^2 \left|w^\prime\right|^2 d\xi^2~,
 &
 w^\prime&=\frac{\partial w}{\partial\xi}~.
\end{align}
In the following we will discuss the action and equations of motion for $(p,q)$ strings in more detail, and derive the conditions for them to end on a pole on the boundary of $\Sigma$.

\subsubsection{Action and equation of motion}
A $(p,q)$ string is a bound state of $q$ D1-branes and $p$ fundamental strings.
From the point of view of the worldvolume gauge theory of $q$ coinciding D1-branes it corresponds to turning on
$p$ units of electric flux \cite{Witten:1995im}. 
Alternatively, one can directly work with an $SL(2,\RR)$ covariant formulation of the string action. This can be realized by introducing two worldvolume gauge fields \cite{Townsend:1997kr,Cederwall:1997ts}, which, however, carry no degrees of freedom and can in turn be integrated out. We will work with the $SL(2,\RR)$ covariant action as given in \cite{Bergshoeff:2006gs}, where the worldvolume gauge fields have been eliminated.\footnote{A detailed account of how the different formulations are related can be found in  \cite{Kluson:2016pxg}.} The string action is then given by
\begin{align}
 S_{(p,q)}&=-T \int d^2\xi \sqrt{\q^T\cM \q}\sqrt{-\det(g)}-T \int \left(p B_2-q C_{(2)}^{\rm RR}\right)~,
 &
 T&=\frac{1}{2\pi\alpha^\prime}~.
\end{align}
$C_{(2)}^{\rm RR}$ and $B_2$ are the pullbacks of the background R-R and NS-NS two-form fields, respectively, and $g$ denotes the pullback of the $SL(2,\RR)$-invariant Einstein-frame metric.
With the dilaton in the conventions of  \cite{DHoker:2016ujz,DHoker:2016ysh,DHoker:2017mds} (see eq.\  (\ref{eq:B-def})), 
\begin{align}\label{eq:cM}
 \q^T\cM \q&=e^{2\phi}\begin{pmatrix}p\\ q\end{pmatrix}^T \begin{pmatrix}
                               1 & -\chi \\ -\chi & \ \chi^2+e^{-4\phi} \ 
                             \end{pmatrix}\begin{pmatrix}p \\ q\end{pmatrix}~.
\end{align}
The action for fundamental strings is recovered for $(p,q)=(\pm 1,0)$, while the action for D1-branes is recovered for $(p,q)=(0,\pm 1)$.

The pullbacks of the NS-NS and R-R background two-form fields to the worldvolume vanish, since they have no components in the time direction. Reduced to the embedding ansatz described above, the action therefore becomes
\begin{align}\label{eq:pq-action}
 S_{(p,q)}&=\int dt d\xi L_{(p,q)}~,
 &
 L_{(p,q)}&=-2Tf_6 \rho|w^\prime|\sqrt{\q^T\cM\q}~.
\end{align}
The resulting equation of motion for the embedding function $w(\xi)$ reads 
\begin{align}\label{eq:pq-eom}
 0&=\frac{\bar w^{\prime\prime}}{\bar w^\prime}-\frac{w^{\prime\prime}}{w^\prime}+\left(\bar w^\prime\partial_{\bar w}-w^\prime\partial_w\right)\ln\big(f_6^2\rho^2 \q^T\cM\q\big)~.
\end{align}

\subsubsection{Boundary conditions}
Eq.~(\ref{eq:pq-eom}) constrains the embedding of the string inside $\Sigma$. Natural end points are the poles on the boundary of $\Sigma$, corresponding to 5-branes, and punctures in $\Sigma$ corresponding to 7-branes. To determine which poles a string can end on, we analyze the behavior of the on-shell Lagrangian in (\ref{eq:pq-action}) near a pole. From the near-pole behavior of the Einstein-frame metric functions in (\ref{eq:Einst-metric-exp}), we conclude that $f_6^2\rho^2=\mathcal O(r^{-1}|\ln r|^{1/2})$, which is not integrable at $r\rightarrow 0$. A finite action is therefore obtained only if
\begin{align}\label{eq:qMq-r0}
 \lim_{r\rightarrow 0}\q^T\cM\q&=0~.
\end{align}
At a pole with NS5 charge, $e^{-2\phi}$ vanishes as the pole is approached and $e^{2\phi}$ diverges.
Realizing (\ref{eq:qMq-r0}) therefore requires
\begin{align}
 p-q\chi_m&=0~,
\end{align}
where we introduced $\chi_m\equiv\lim_{w\rightarrow r_m}\chi$. With (\ref{eq:axion-dilaton-near-pole}) and (\ref{eq:residue-N}), $\chi_m$ evaluates to
\begin{align}
 \chi_m&=\frac{N_{\rm D5}}{N_{\rm NS5}}~.
\end{align}
Near a D5-pole, $e^{-2\phi}$ in (\ref{eq:chi-phi-D5}) diverges as $r\rightarrow 0$, and  $\q^T\cM\q$ remains finite only if $q=0$.
In order for a $(p,q)$ string ending on a generic $(P,Q)$ 5-brane pole to have finite action, the string charges therefore have to be related to the 5-brane charges by
\begin{align}
Pq-pQ&=0~.
\end{align}
This is as expected from string theory.
In particular, only fundamental strings can end on D5-brane poles and only D1-branes can end on NS5-brane poles.

\subsubsection{Scaling dimension and charge}\label{sec:scaling-charge}
The scaling dimension of the dual operator for a given $(p,q)$ string embedding is given by the on-shell Hamiltonian, and can be obtained here as
\begin{align}\label{eq:pq-Delta}
 \Delta_{(p,q)}=\int d\xi L_{(p,q)}~.
\end{align}
The $(p,q)$ string also couples to the background two-form field. Although this coupling vanishes for the background solution, due to the specific form of the embedding, the coupling to the fluctuation $\delta C_{(2)}$ discussed in sec.~\ref{sec:gauge-field} is non-trivial. 
The coupling is
\begin{align}
 \delta S_{(p,q)}&=
 T \int \left(p \delta B_2-q \delta C_{(2)}^{\rm RR} \right)
 \nonumber\\
 &=T \int \left(p\Re(d\cC)-q\Im(d\cC)\right)\wedge f_I A^I~.
\end{align}
Turning on the $I=3$ Cartan direction, for which $f_I=\cos\theta_1$, and choosing the highest weight state in the multiplet corresponding to locating the string at $\theta_1=0$, we can identify the R-charge as
\begin{align}\label{eq:pq-Q}
 Q_{(p,q)}&=T\int_{\Sigma_{(p,q)}} \left(p\Re(d\cC)-q\Im(d\cC)\right)~,
\end{align}
where $\Sigma_{(p,q)}$ denotes the cycle that the string wraps in $\Sigma$.
For string embeddings that do not cross branch cuts, this becomes the difference in $T\left(p\Re(\cC)-q\Im(\cC)\right)$ between the two end points of the string.

With the expressions for scaling dimension and $R$-charge in hand, we will be able to explicitly verify that the BPS relation for the supersymmetric states of interest here, which is \cite{Minwalla:1997ka,Bhattacharya:2008zy}
\begin{align}\label{eqn:BPS-rel}
 \Delta&=3Q~,
\end{align}
is satisfied for the strings and string webs to be discussed in the next sections.

\subsection{The \texorpdfstring{$+_{N,M}$ and $X_{N,M}$}{+-N-M and X-N-M} solutions}\label{sec:grid-solution}

Let us begin with the solution dual to the $+_{N,M}$ theory, originally introduced in sec.~4.2 of \cite{DHoker:2017mds}.
The poles on the real line are placed at
\begin{align}
 r_1&=1~,
 &
 r_2&=\frac{2}{3}~,
 &
 r_3&=\frac{1}{2}~,
 &
 r_4&=0~,
\end{align}
and the regularity conditions fix $\cA_+^0=Z_+^2\ln 3-Z_+^1\ln 2$. The parameters $s_1$, $s_2$ are chosen as the two solutions to the quadratic equation
\begin{align}
 M\left(4 s^2-6 s+2\right)+i N (2-3 s) s&=0~,
\end{align}
and are both in the upper half plane. Finally, the overall normalization of the residues, $\sigma$, is chosen as
\begin{align}
 \sigma&=\frac{4 M - 3 iN}{8}\alpha^\prime~.
\end{align}
The residues produced by this choice of parameters are pairwise opposite equal and given by
\begin{align}\label{eq:D5-NS5-residues}
 -Z_+^1=Z_+^3&=\frac{3}{4}i \alpha^\prime N~,&
 Z_+^2=-Z_+^4&=\frac{3}{4}\alpha^\prime M~.
\end{align}
With the results of sec.~\ref{sec:asympt-charge}, these are the appropriate residues for an intersection of $N$ D5-branes  and $M$ NS5-branes, and this solution is thus expected to be the holographic dual for the $+_{N,M}$ theory discussed in sec.~\ref{sec:grid-N-M}.

\begin{figure}
\centering
\includegraphics{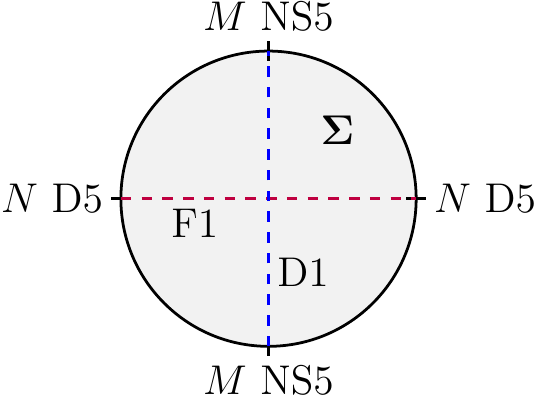}
\caption{Disc representation of the solution with residues given in (\ref{eq:D5-NS5-residues}) along with the embeddings of a 
D1-brane and a fundamental string.
\label{fig:disc}}
\end{figure}

The solution has two $\ZZ_2$ symmetries, which will be instrumental in discussing the string embeddings. Their action takes a simple form after mapping the upper half plane to the unit disc centered at the origin, such that the poles are on the intersections of the boundary of the disc with the real and imaginary axes. Combining an $SL(2,\RR)$ transformation on the upper half plane, mapping three of the poles to $\lbrace-1, 0, 1\rbrace$ and the remaining one to infinity, with a Cayley transform, yields
\begin{align}
 w&=\frac{f(z)+1}{f(z)+2}~,
 &
 f(z)&=i\frac{1+z}{1-z}~,
\end{align}
where $w$ is the complex coordinate on the upper half plane and $z$ the coordinate on the disc.
This leads to the solution in the form illustrated in fig.~\ref{fig:disc}.
The poles are mapped to the boundary of the disc as follows
\begin{align}
 r_1&\rightarrow z=1~,
 &
 r_2&\rightarrow z=-i~,
 &
 r_3&\rightarrow z=-1~,
 &
 r_4&\rightarrow z=i~.
\end{align}
When formulated in the $z$ coordinate on the disc, the supergravity fields transform as follows under reflection across the real and imaginary axes
\begin{align}\label{eq:D5NS5-symmetry}
 z&\rightarrow \pm\bar z: &
 (f_6^2,f_2^2,\rho^2,\tau,\cC-\cC_0)&\rightarrow (f_6^2,f_2^2,\rho^2,-\bar\tau,\pm (\bar\cC-\bar\cC_0))~,
\end{align}
where $\cC_0$ is the value of $\cC$ at the center of the disc, $\cC_0=\cC\vert_{z=0}$.

\subsubsection{String embeddings}

As discussed in sec.~\ref{sec:D1F1}, fundamental strings can connect D5-brane poles with finite action, while D1-branes can connect NS5-brane poles with finite action. We thus expect to find these two embeddings in the supergravity solution corresponding to a D5/NS5 brane intersection. The embeddings are expected to take a particularly simple form in the $z$ coordinate where $\Sigma$ corresponds to a disc and the $\ZZ_2$ symmetries are transparent.

For a fundamental string with $(p,q)=(1,0)$, 
we expect the embedding to correspond to the segment of the real axis connecting the poles on the real line in fig.~\ref{fig:disc}. The equation of motion (\ref{eq:pq-eom}) for such an embedding with $\bar z=z=\xi$ turns into
\begin{align}
 0&=\left(\partial_{\bar z}-\partial_z\right)\ln\big(e^{2\phi} f_6^2 \rho^2\big)~.
\end{align}
Due to the transformation of the supergravity fields under reflection across the real line as given in (\ref{eq:D5NS5-symmetry}), the derivative acts on a term which is odd under reflection across the real line, and therefore vanishes. The anticipated embedding therefore indeed satisfies the equation of motion.
In the $w$ coordinates the embedding corresponds to a half circle in the upper half plane connecting $r_1$ and $r_3$, namely,
\begin{align}
 w_{\rm F1}&=\frac{1}{4}\left(3+e^{i\xi}\right)~,
 &
 \xi&\in [0,\pi]~.
\end{align}
This is the supergravity realization of the fundamental string connecting D7-branes in fig.~\ref{fig:grid-1}.

By S-duality, one expects to find a D1-brane which takes the form of a straight line connecting the NS5-poles in fig.~\ref{fig:disc} along the imaginary axis.
The equation of motion (\ref{eq:pq-eom}), with $z=i\xi$ and $(p,q)=(0,1)$, evaluates to
\begin{align}
  \left(\partial_{\bar z}+\partial_z\right)\ln\left(f_6^2 \rho^2 \tau\bar \tau\right)&=0~.
\end{align}
Again, due to the transformation of the supergravity fields under reflection across the imaginary axis, as given in (\ref{eq:D5NS5-symmetry}), the derivative acts on a function which is odd under reflection across the imaginary axis, and vanishes. The embedding therefore again solves the equation of motion.
In the upper half plane the solution for the D1-brane embedding is a half circle connecting $r_2$ and $r_4$,
\begin{align}
 w_{\rm D1}&=\frac{1}{3}\left(1+e^{i\xi}\right)~, & \xi &\in [0,\pi]~.
\end{align}
This provides the supergravity realization of the D1-brane connecting $(0,1)$ 7-branes in fig.~\ref{fig:grid-1}.

To determine the scaling dimension and $R$-charge of the string states, we realized the background solution for a large number of explicit choices for $N$ and $M$, and evaluated (\ref{eq:pq-Delta}) and  (\ref{eq:pq-Q}) numerically. We found striking agreement,  up to machine precision of $\mathcal O(10^{-16})$, with simple analytic formulas. The results for the scaling dimensions and $R$-charges are
\begin{align}\label{eq:Delta-Q-D5NS5}
 \Delta_{\rm F1}&=\frac{3}{2}M~,
 &
 Q_{\rm F1}&=\frac{1}{3}\Delta_{\rm F1}~,
 \nonumber\\
  \Delta_{\rm D1}&=\frac{3}{2}N~,
 &
 Q_{\rm D1}&=\frac{1}{3}\Delta_{\rm D1}~.
\end{align}
In particular, scaling dimension and charge for the fundamental string are independent of $N$ and depend linearly on the number of NS5-branes, while the results for the D1-brane are independent of $M$ and depend linearly on the number of D5-branes.
The scaling dimensions, including their $N$ and $M$ dependence as well as numerical coefficients, agree precisely with those obtained from the field theory discussion in sec.~\ref{sec:grid-N-M}, and summarized in tab.~\ref{tab:CFT1}, in the limit of large $N$ and $M$.

\begin{figure}
\center
\includegraphics[height=0.25\textwidth]{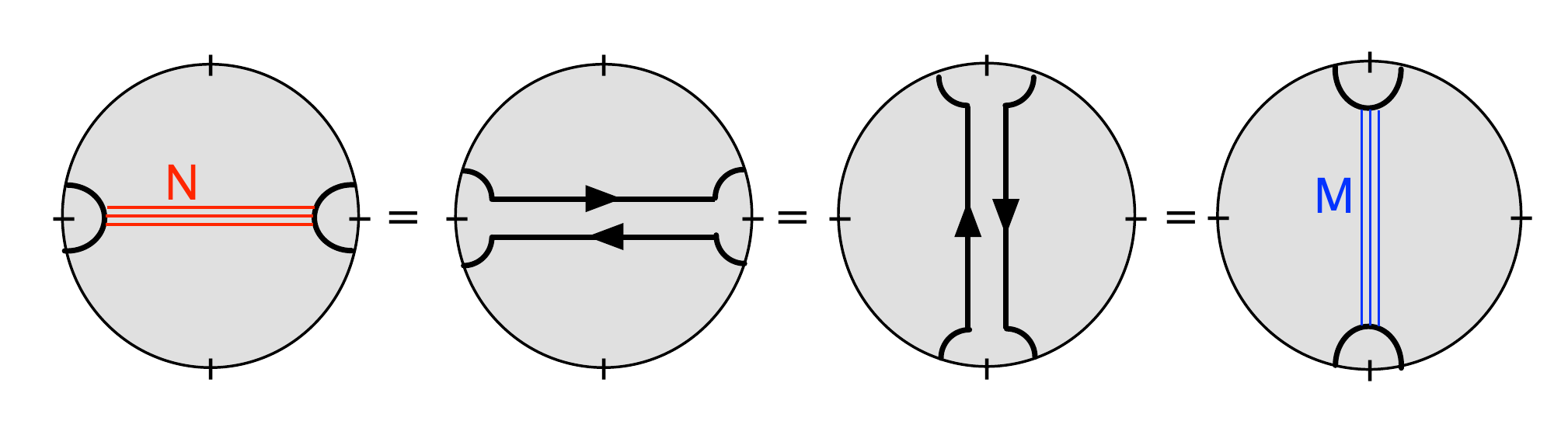} 
\caption{The chiral ring relation as realized in the dual geometry: $N$ fundamental strings between the D5-brane poles are
equivalent to a D3-brane wrapping the sum of the two D5-brane cycles, which is equivalent to a D3-brane wrapping the
sum of the two NS5-brane cycles, which in turn is equivalent to $M$ D1-branes between the NS5-brane poles.}
\label{ChiralRing}
\end{figure}

Finally, the global $U(1)$ charge of these states can be determined as follows.
Recall that the global $U(1)$ symmetries correspond to the flux-free combinations of 3-cycles surrounding the poles.
In this case there is only one, corresponding to the sum of the $(1,0)$ and $(-1,0)$ cycles, or equivalently to the
sum of the $(0,1)$ and $(0,-1)$ cycles.
A D3-brane wrapping either of these cycles describes a particle charged under the $U(1)$ symmetry.
In the first case it is equivalent to the combination of a D3-brane on the $(1,0)$ cycle and a D3-brane on the $(-1,0)$ cycle 
connected by $N$ fundamental strings, as required by tadpole cancellation on both 3-branes,
and in the second case to D3-branes on the $(0,1)$ and $(0,-1)$ cycles connected by $M$ D1-branes. 
The equivalence of the two combinations is naturally interpreted as the geometrical description of the chiral ring relation (\ref{eq:chiral-ring}),
fig.~\ref{ChiralRing}.
We can also conclude from this, with a suitable choice of normalization, that the single fundamental string state carries $M$ units of $U(1)$ charge,
and the single D1-brane state carries $N$ units of $U(1)$ charge, as seen in sec.~\ref{sec:grid-N-M}.

\subsubsection{The \texorpdfstring{$X_{N,M}$}{X} solution}

The supergravity solution dual to the $X_{N,M}$ theory is closely related to the dual of the $+_{N,M}$ theory.
It is realized by the same choice of parameters that was used for the $+_{N,M}$ solution, 
except for a different choice of $\sigma$, which is now given by
\begin{align}
 \sigma&=\frac{1+i}{8} \alpha^\prime(4 M-3iN)~.
\end{align}
This realizes the residues for an intersection of $N$ $(1,-1)$ 5-branes and $M$ $(1,1)$ 5-branes,
\begin{align}
 Z_+^1&=-Z_+^3=\frac{3}{4}\alpha^\prime(1-i)N~,
 &
 Z_+^2&=-Z_+^4=\frac{3}{4}\alpha^\prime(1+i)M~.
\end{align}
We now have $(1,-1)$ strings connecting the $(1,-1)$ 5-brane poles at $r_1$ and $r_3$, and $(1,1)$ strings connecting the $(1,1)$ 5-brane poles at $r_2$ and $r_4$. 
The results for the scaling dimensions and charges are
\begin{align}\label{eq:11-1m1-DeltaQ}
 \Delta_{(1,-1)}&=3M~,
 &
 Q_{(1,-1)}&=\frac{1}{3}\Delta_{(1,-1)}~,
 \nonumber\\
  \Delta_{(1,1)}&=3N~,
 &
 Q_{(1,1)}&=\frac{1}{3}\Delta_{(1,1)}~.
\end{align}
These results are related to the ones for the D5/NS5 intersection in (\ref{eq:Delta-Q-D5NS5}) by a simple rescaling by a factor $2$.
This can be understood from the supergravity perspective as follows: The solution realizing an intersection of $N$ D5 and $M$ NS5 branes can be related by an $SL(2,\RR)$ transformation to an intersection of $N/\sqrt{2}$ $(1,-1)$ 5-branes and $M/\sqrt{2}$ $(1,1)$ 5-branes. Likewise, the configuration with $N$ D5 branes, $M$ NS5 branes and a fundamental string or a D1-brane is related by $ SL(2,\RR)$ to a configuration with $N/\sqrt{2}$ $(1,-1)$ 5-branes, $M/\sqrt{2}$ $(1,1)$ 5-branes and a $(1,-1)/\sqrt{2}$ or $(1,1)/\sqrt{2}$ string, respectively. For this configuration, the scaling dimensions of the string states are still given, respectively, by $3/2N$ and $3/2M$. Upon rescaling $M$ and $N$, as well as the string charges, by factors of $\sqrt{2}$, to realize the $X_{N,M}$ solution with a $(1,-1)$ or $(1,1)$ string, we recover the result in (\ref{eq:11-1m1-DeltaQ}).

We emphasize that this reasoning is justified in the supergravity approximation, where the $SL(2,\ZZ)$ duality of Type IIB string theory is enhanced to $SL(2,\RR)$. This corresponds to the ``large-$N$'' limits of the $+_{N,M}$ and $X_{N,M}$ theories. In the string theory description, where the 5-brane charges are quantized and the S-duality group is reduced to $SL(2,\ZZ)$, the corresponding brane webs are not related by S-duality.

As in the $+_{N,M}$ solution, we can determine the global $U(1)$ charge of these states by considering 
a D3-brane wrapping a flux-free combination of 3-cycles.
This shows that the $(1,-1)$ string carries $M$ units of charge and the $(1,1)$ string carries $N$ units of charge.
As before this also provides a geometrical description of a chiral ring-like relation analogous to (\ref{eq:chiral-ring}).
However, in the case of the $X_{N,M}$ theory we do not have an explicit realization of this relation in terms
of operators in the quiver gauge theory, since we are not able to construct both operators simultaneously
in a given gauge theory.

\subsection{The \texorpdfstring{$T_N$ and $Y_N$}{T-N and Y-N} solutions}\label{sec:T-N-sol}
\begin{figure}
\centering
\includegraphics{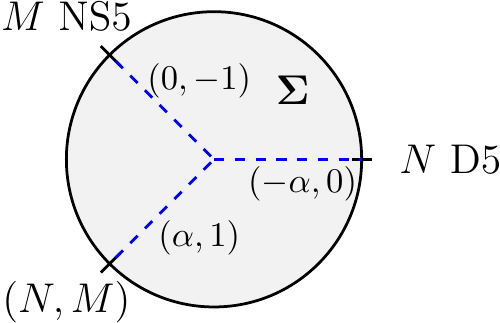}
\caption{Disc representation of the 3-pole solutions with residues given in (\ref{eq:N-junction-residues}).
The dashed blue line shows a string junction connecting all three poles, with $\alpha=N/M$.\label{fig:njunction-disc}}
\end{figure}

In this section we realize the supergravity solution corresponding to the $T_N$ theory.
This solution was originally introduced in an $SL(2,\RR)$ transformed version in \cite{DHoker:2017mds}. We start with a slightly more general charge assignment and specialize to the $T_N$ case at the end. 
The solution has three poles, $L=3$, which are located at
\begin{align}\label{eq:3-poles}
 r_1&=1~, & r_2&=0~, & r_3=-1~.
\end{align}
The regularity conditions are solved by setting $\cA_+^0=\sigma s_1 \ln 2$, and the remaining parameters are chosen as
\begin{align}
 s_1&=\frac{iN}{iN +2M}~,
 &
 \sigma&=\frac{3}{4}\alpha^\prime \frac{iN}{s_1}~.
\end{align}
This realizes the residues for a junction of $M$ NS5 branes and $N$ D5 branes,
\begin{align}\label{eq:N-junction-residues}
 Z_+^1&=\frac{3}{4}\alpha^\prime M~,
 &Z_+^2&=\frac{3}{4}i\alpha^\prime N~,
 &Z_+^3&=-\frac{3}{4}\alpha^\prime(M+iN)~.
\end{align}
This solution has three $\ZZ_2$ symmetries, which are again transparent on the disc. We map the upper half plane with coordinate $w$ to the unit disc centered at the origin with coordinate $z$ via
\begin{align}
w&=\frac{i}{\sqrt{3}}\frac{1-z}{1+z}~.
\end{align}
The solution then takes the form illustrated in fig.~\ref{fig:njunction-disc}, with the poles at the cubic roots of $1$.
There are three distinguished lines, for which the supergravity fields have simple transformations under reflection,
and these are the diameters from a given pole to the diametrically opposed point. 
For reflection across the real line we have
\begin{align}\label{eq:N-junction-symmetry}
 z&\rightarrow \bar z: &
 \left(f_6^2,f_2^2,\rho^2,\tau-\chi_{2}\right)&\rightarrow 
 \left(f_6^2,f_2^2,\rho^2,\chi_{2}-\bar\tau\right)~,
\end{align}
where $\chi_{\ell}$ denotes the asymptotic value of $\chi$ at the pole $r_\ell$.
This symmetry can be understood as follows: We can transform the charge assignment in (\ref{eq:N-junction-residues}) to a configuration where $Z_+^1=-\overline{Z_+^3}$ by an $SL(2,\RR)$ transformation which leaves $Z_+^2$, i.e.\ the residue at the D5-brane pole, invariant. This is a D7-brane monodromy transformation and simply produces a shift in the axion. The configuration with $Z_+^1=-\overline{Z_+^3}$ now has a manifest reflection symmetry, which is analogous to $z\rightarrow\bar z$ in (\ref{eq:D5NS5-symmetry}). Applying the inverse $SL(2,\RR)$ transformation, back to the residues in (\ref{eq:N-junction-residues}), produces the transformation of the supergravity fields under reflection as  given in (\ref{eq:N-junction-symmetry}).
By analogous reasoning, one finds that the Einstein-frame metric functions are invariant under $z\rightarrow e^{4\pi i/3}\bar z$ and $z\rightarrow e^{8\pi i/3}\bar z$, while the axion-dilaton scalar changes by an $SL(2,\RR)$ transformation.

\subsubsection{String embeddings}

The transformations in (\ref{eq:N-junction-symmetry}) are sufficient to show that the equation of motion for a
fundamental string is satisfied along $z\in\RR$, which in the $w$ coordinate on the upper half plane corresponds to the imaginary axis
\begin{align}
w_{\rm F1}&=\frac{i\xi}{1-\xi}~,&
\xi&\in(0,1)~.
\end{align}
Likewise, the two further $\ZZ_2$ symmetries discussed above suggest that there are also solutions for D1-branes connecting the NS5 pole $r_1$ to the diametrically opposed point along the diameter in the $z$ coordinate, and $(N,M)$ strings connecting the $(N,M)$ 5-brane at $r_3$ to the diametrically opposed point on the boundary of $\Sigma$. Mapping back to $w$, this yields the following additional embeddings
\begin{align}
 w_{\rm D1}&=\frac{1-2e^{-i\xi}}{3}~,
 &
 w_{(N,M)}&=\frac{2e^{i\xi}-1}{3}~,&
 \xi\in(0,\pi)~.
\end{align}
They indeed satisfy the equation of motion (\ref{eq:pq-eom}) with the appropriate charge assignments.

A natural object to consider in order to implement the appropriate boundary conditions for open strings is a string junction, formed out of the segment of each of these strings connecting the pole to the center of the disc. This yields a string junction connecting all three poles, as illustrated in fig.~\ref{fig:njunction-disc}, with $\alpha=N/M$. Note that the $(p,q)$ string charges are conserved at the trivalent vertex.
In the supergravity description one can realize e.g.\ a $(-\alpha,0)$ string with generic $\alpha$; for a full string theory description the constraints from charge quantization have to be taken into account. This will be implemented automatically when we specialize to $N=M$ shortly.
Adding the contribution of the various segments of the string junction to the Hamiltonian, we find the scaling dimension and charge as follows,
\begin{align}\label{eq:T-N-junction}
 \Delta_{(-\alpha,0)-(0,-1)-(\alpha,1)}&=\frac{3}{2}N~,
 &
 Q_{(-\alpha,0)-(0,-1)-(\alpha,1)}&=\frac{1}{3}\Delta_{(-\alpha,0)-(0,-1)-(\alpha,1)}~.
\end{align}
For $M=N$, the background solution describes a junction of $N$ D5-branes and $N$ NS5-branes, and realizes the supergravity dual for the $T_N$ theory discussed in sec.~\ref{sec:T-N}. With $\alpha=1$ in that case, the string junction joins a D1-brane and a fundamental string with a $(1,1)$ string, with charge and scaling dimension given in (\ref{eq:T-N-junction}).
The string junction is the supergravity realization of the junction shown in fig.~\ref{TNjunction}(a), and the scaling dimension precisely agrees with that of the trifundamental operator discussed in sec.~\ref{sec:T-N} in the large-$N$ limit.

It would be interesting to generalize this configuration to the general $k$ case, whose scaling dimension is $\Delta=\frac{3}{2}\,k\,(N-k)$. Note that this is symmetric under the exchange $k\leftrightarrow N-k$. Hence, it is natural to expect this configuration to be similar to a $k$-string.

\subsubsection{The \texorpdfstring{$Y_N$}{Y-N} solution}

By analogy with the discussion for the $+_{N,M}$ and $X_{N,M}$ solutions, the $Y_N$ solution can be obtained
from the $T_N$ solution by an $SL(2,\RR)$ transformation combined with a charge rescaling, and the corresponding field theories are related at large $N$. More specifically, the supergravity solution for a junction of $N$ D5-branes and $N$ NS5-branes with $N$ $(1,1)$ 5-branes is related by $SL(2,\RR)$ to the supergravity solution for a junction of $N/\sqrt{2}$ $(1,1)$ 5-branes and $N/\sqrt{2}$ $(-1,1)$ 5-branes with $N/\sqrt{2}$ $(0,-2)$ 5-branes. The string junction describing the BPS state in the $T_N$ solution, which was  joining a fundamental
string and a D1-brane with a $(1,1)$ string, is mapped by this $SL(2,\RR)$ transformation to a junction of a $(1,1)/\sqrt{2}$ string and a $(-1,1)/\sqrt{2}$ string with a $(0,-2)/\sqrt{2}$ string. Rescaling the 5-brane charges and the string charges by factors of $\sqrt{2}$ in order to realize the $Y_N$ solution with a junction of $(1,1)$, $(-1,1)$ and $(0,-2)$ strings again produces a factor $2$. The resulting scaling dimension and charge are
\begin{align}
 \Delta_{(1,1)-(-1,1)-(0,-2)}&=3N~,
 &
 Q_{(1,1)-(-1,1)-(0,-2)}&=\frac{1}{3}\Delta_{(1,1)-(-1,1)-(0,-2)}~.
\end{align}
This supergravity result for the string web shown in fig.~\ref{YNjunction}(a) precisely agrees with the scaling dimension of the  $(({\bf 2N})^2_{asym},{\bf N},{\bf N})$ operators discussed in sec.~\ref{sec:Y-N} at large $N$.

\subsection{The \texorpdfstring{$+_{N,M,k}$}{monodromy} solution}

\begin{figure}
\centering
\includegraphics{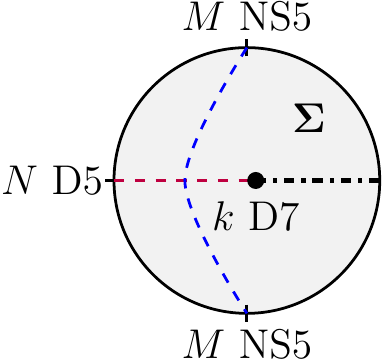}
\caption{Disc representation of the D5/NS5$^2$/D7 supergravity solution, with the black dot-dashed line showing the branch cut. The fundamental string connecting the D5-brane pole to the D7 brane puncture is shown as purple dashed line, the D1-brane connecting the NS5-brane poles is shown as blue dashed line. \label{fig:D5-NS52-D7}}
\end{figure}

We now turn to a configuration with D7-brane monodromy, and identify string probes in the setup engineered in sec.~4.3 of \cite{Gutperle:2018vdd}. It has three poles, and the parameters associated with the seed solution are chosen as
\begin{align}
 r_1&=1~, & r_2&=0~, & r_3&=-1~,
 &
 \sigma&=\frac{3}{2}\alpha^\prime M~,
 &s_1&=\frac{i N}{2M}~.
\end{align}
The location of the branch point, $w_1$, the phase $\gamma_1$ fixing the orientation of the associated branch cut, and the strength of the monodromy $n_1^2$ are chosen as
\begin{align}
 n_1^2&=k~, &
 w_1&=i\lambda~,
 &
 \lambda&=\cot\left(\frac{\pi N}{2M k}\right)~,
 &
 \gamma&=-1~,
\end{align}
with $k>0$.
This satisfies the regularity conditions and realizes the residues
\begin{align}
 \cY_+^1&=\frac{4}{3}\alpha^\prime M~,
 &
 \cY_+^2&=\frac{4}{3}\alpha^\prime iN~,
 &
 \cY_+^3&=-\frac{4}{3}\alpha^\prime M~.
\end{align}
This produces a solution with one external stack of $N$ D5-branes, two external stacks of $M$ NS5-branes, and $k$ D7-branes at $w_1$.  The supergravity solution is illustrated in fig.~\ref{fig:D5-NS52-D7}. It realizes the dual for the $+_{N,M,k}$ theory discussed in sec.~\ref{sec:perp}, and the corresponding brane webs are illustrated in fig.~\ref{GridNMk}.

In this background we expect to find a fundamental string connecting the D5-brane pole to the D7-brane puncture, as well as a D1-brane connecting the two NS5-brane poles. For the fundamental string connecting the D5-brane pole to the D7 branes, the embedding can be inferred from a $\ZZ_2$ symmetry of the setup, which acts as reflection across the imaginary axis in the disc representation of fig.~\ref{fig:D5-NS52-D7}. The embedding of the 
fundamental string into the upper half plane is given by
\begin{align}
 w_{\rm F1}&=i\xi~, & \xi&\in(0,\lambda)~,
\end{align}
and indeed solves the equation of motion. The scaling dimension and charge are 
\begin{align}
 \Delta_{\rm F1}&=\frac{3}{2}\left(M-\frac{N}{k}\right)~,
 &
 Q_{\rm F1}&=\frac{1}{3}\Delta_{\rm F1}~.
\end{align}
This realizes the $({\bf N},{\bf \bar k},{\bf 1},{\bf 1})$ operator of tab.~\ref{tab:CFT1} and the scaling dimension agrees with the field theory arguments of sec.~\ref{sec:perp} at large $M$ and $N/k$.

For the D1-brane there is no simple symmetry argument fixing the embedding. However, the qualitative form of the embedding can be inferred,  and that qualitative form of the embedding is sufficient to deduce the scaling dimension of the dual operator. To argue for the form of the embedding, we start with the limit where $k$ is large, and the puncture approaches the boundary of the disc at the point diametrically opposed to the D5-brane pole in fig.~\ref{fig:D5-NS52-D7}. As discussed in more detail in sec.~4.4 of \cite{Gutperle:2018vdd}, the solution reduces to a four-pole solution without monodromy in that limit. For this four-pole solution we found the embedding of the 
D1-brane in sec.~\ref{sec:grid-solution}, and it corresponds to a vertical line connecting the NS5-brane poles. As the puncture is moved inwards along the equator of the disc, decreasing $k$, the embedding gets deformed, to qualitatively take the form illustrated in fig.~\ref{fig:D5-NS52-D7}.
The scaling dimension can be inferred from the $R$-charge, which can be computed from the values of the background two-form field at the end points of the embedding, as discussed in sec.~\ref{sec:scaling-charge}. This yields
\begin{align}
 Q_{\rm D1}&=\frac{1}{2}N~.
\end{align}
Using the BPS relation (\ref{eqn:BPS-rel}), we then conclude that $\Delta_{\rm D1}=\frac{3}{2}N$, in agreement with the value inferred from field theory considerations in sec.~\ref{sec:perp}.

\subsection{The \texorpdfstring{$\pslash_N$}{pslash-N} solution}

We now realize an intersection with six external 5-brane stacks, as discussed in sec.~\ref{sec:pslash}, corresponding to a six-pole supergravity solution with residues
\begin{align}\label{eq:6pole-residues}
 Z_+^1&=i Z_+^2=-Z_+^4=-iZ_+^5=\frac{3}{4}\alpha^\prime N~,
 &
 -Z_+^3&=Z_+^6=\frac{3}{4}\alpha^\prime (1+i)N~.
\end{align}
To exploit the symmetries of this setup, we start the construction directly on the disc.
We introduce a coordinate on the disc, $z$, a mapping from the disc to the upper half plane, $f$, and place the poles as follows,
\begin{align}
 r_\ell&=f(e^{\frac{i\pi\ell}{3}})~,
 &
 f(z)&=\frac{i-z}{1-iz}~.
\end{align}
On the disc we expect a simple transformation of the configuration under discrete transformations mapping the poles and residues into each other, in particular under $z\rightarrow -z$ and $z\rightarrow \bar z$. 
We recall that the $\lbrace s_n\rbrace$ have an interpretation as the locations of auxiliary charges for a certain electrostatics potential in the construction of \cite{DHoker:2017mds}. They should be mapped into themselves under the discrete symmetry transformations. To realize the residues in (\ref{eq:6pole-residues}) and solve the regularity conditions in (\ref{eqn:constr}), we choose 
\begin{align}
 s_n&=f\left(-\sqrt[4]{\sqrt{3}-2}\,e^{in\pi/2}\right)~,
 &
 n&=1,\cdots,4~,
\end{align}
while the overall normalization $\sigma$ and constant $\cA_+^0$ are fixed as
\begin{align}
 \sigma&=-\frac{9}{2}(1+i)\alpha^\prime N~,
 &\cA_+^0&=\frac{3}{8}(1-i)\alpha^\prime N \ln(7+4\sqrt{3})~.
\end{align}

\begin{figure}
\centering
\subfigure[][]{\label{fig:6-pole-1}
\includegraphics{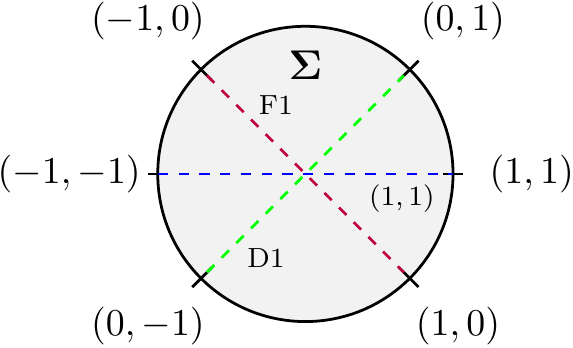}
}\hskip 0.6in
\subfigure[][]{\label{fig:6-pole-2}
\includegraphics{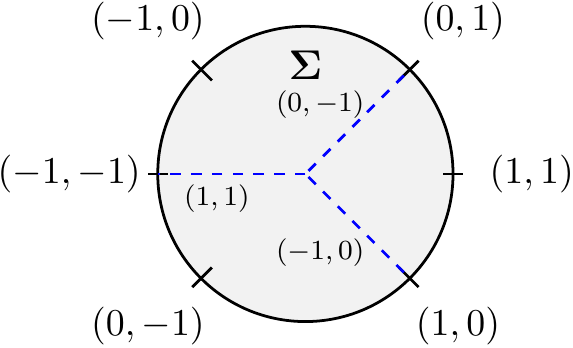}
}
\caption{Disc representation of the 6-pole solutions with residues given in (\ref{eq:6pole-residues}).
On the left hand side strings connecting like poles. On the right hand side the three-junction.
\label{fig:6pole-disc}}
\end{figure}

There are various embeddings of strings and string junctions into these solutions. There are three strings connecting poles with opposite-equal 5-brane charges, as shown in fig.~\ref{fig:6-pole-1}. The scaling dimensions and $R$-charges are given by
\begin{align}
 \Delta_{\rm F1}&=3N~,
 &
 \Delta_{\rm D1}&=3N~,
 &
 \Delta_{(1,1)}&=3N~,
 \nonumber\\
 Q_{\rm F1}&=\frac{1}{3}\Delta_{\rm F1}~,
 &
 Q_{\rm D1}&=\frac{1}{3}\Delta_{\rm D1}~,
 &
 Q_{(1,1)}&=\frac{1}{3}\Delta_{(1,1)}~.
\end{align}
This agrees precisely with the scaling dimensions derived in sec.~\ref{sec:pslash} and summarized in tab.~\ref{tab:CFT1}.
Using segments of the strings, one can also form 3-pronged string junctions connecting three poles. 
With a $(1,1)$ string connecting the $(-1,-1)$ 5-branes at $r_3$ to the center of the disc, a $(-1,0)$ string connecting the NS5-branes at $r_1$ to the center and a $(0,-1)$ string connecting the D5-branes at $r_5$ to the center of the disc, one forms the junction shown in fig.~\ref{fig:6-pole-2}. The scaling dimension and $R$-charge are
\begin{align}
 \Delta_{(1,1)-(-1,0)-(0,-1)}&=\frac{9}{2}N~,
 &
 Q_{(1,1)-(-1,0)-(0,-1)}&=\frac{1}{3}\Delta_{(1,1)-(-1,0)-(0,-1)}~.
\end{align}
An analogous junction can be formed to connect the remaining three poles, with the same result for the scaling dimension. These results agree with the results of sec.~\ref{sec:pslash} at large $N$.

\begin{acknowledgments}

D.R-G. thanks the Galileo Galilei Institute for Theoretical Physics for hospitality and INFN for partial support during the completion of this work. We also thank the Aspen Center for Physics and the organizers of the 2017 winter conference on Superconformal Field Theories in Four or More Dimensions. O.B. is supported in part by the Israel Science Foundation under grant no. 352/13, and by the US-Israel Binational Science Foundation under grant no. 2012-041. CFU is supported in part by the National Science Foundation under grant PHY-16-19926.  D.R-G is partially supported by the Asturias Government grant FC-15-GRUPIN14-108 and Spanish Government grant MINECO-16-FPA2015-63667-P.
\end{acknowledgments}

\bibliographystyle{JHEP}
\bibliography{ads6}
\end{document}